\documentclass[12pt,preprint]{aastex}
\usepackage{emulateapj5}

\def\TkeV#1{kT\ =\ #1\ keV~}
\def\TkeVsim#1{kT\ $\sim$#1\ keV~}

\shorttitle{Smoothed Particle Inference}
\shortauthors{Peterson, Andersson \& Marshall}

\begin{document}

\title{Smoothed Particle Inference: \\A Kilo-Parametric Method for X-ray Galaxy Cluster Modeling}


\author{J.R.~Peterson$^1$, P.J.~Marshall$^1$, K.~Andersson$^{2,3}$}

\affil{$^1$ KIPAC, Stanford University, PO Box 20450, Stanford, CA 94309, USA}

\email{jrpeters@slac.stanford.edu, pjm@slac.stanford.edu}

\affil{$^2$ Department of Physics, Stockholm University, Albanova University Center, S-106 91, Stockholm, Sweden}

\affil{$^3$ Stanford Linear Accelerator Center, Menlo Park, CA 94025, USA}

\email{kanderss@slac.stanford.edu}


\begin{abstract}  

We propose an ambitious new method that models the intracluster medium
in clusters of galaxies as a set of X-ray emitting smoothed particles of
plasma. Each smoothed particle is described by a handful of parameters
including  temperature, location, size, and elemental abundances.
Hundreds to thousands of these particles are used to construct a model
cluster of galaxies, with the appropriate complexity estimated from the
data quality. This model is then compared iteratively with X-ray data in
the form of adaptively binned photon lists via a two-sample likelihood
statistic and iterated via Markov Chain Monte Carlo. The complex cluster
model is propagated through the X-ray instrument response using direct
sampling Monte Carlo methods. Using this approach the method can
reproduce many of the features observed in the X-ray emission in a less
assumption-dependent way than traditional analyses, and it allows for a
more detailed characterization of the density, temperature, and metal
abundance structure of clusters.  Multi-instrument X-ray analyses and
simultaneous X-ray, Sunyaev-Zeldovich (SZ), and lensing analyses are a
straight-forward extension of this methodology.  Significant challenges
still exist in understanding the degeneracy in these models and the
statistical noise induced by the complexity of the models.  

\end{abstract}


\section{Motivation}

There are a large class of data analysis problems in astrophysics
involving a model of a source that has a spectrum that varies spatially
on the sky. In the case of the X-ray emission from clusters of galaxies,
where the individual photon energies are recorded, the problem is
especially prominent.  In fact, much of the recent work in the X-ray
analysis of clusters of galaxies has had the aim of obtaining a complete
description of the density and temperature structure of the intracluster
medium.  

The X-ray emission from clusters of galaxies has been recently shown,
using observations with the Chandra and XMM-Newton telescopes, to have a
very complex distribution.  In general, the surface brightness lacks
circular symmetry. For example, cold fronts, or cluster gas that
``sloshing'' relative to the dark matter gravitational potential, is a
significant perturbation from a spherically-symmetric gas distribution
\citep{markevitch}.  ``Radio bubbles'', which are displacements of
cluster gas by relativistic radio-emitting plasma, have complex
morphologies and vary from cluster to cluster.  \citet{sanders} have
identified a ``swirl'' in the temperature structure of the Perseus
cluster possibly due to rotation of the cluster gas.  There may also be
ripples or waves in the cluster gas \citep{fabian}. Neither are clusters
isothermal: \citet{kaastra} and \citet{buote} have questioned the
assumption that the cluster gas can even approximately be described by a
single phase as a function of radius.   In fact, they find evidence that
the temperature can vary at a single radius by a larger factor than the
average temperature drops radially across the entire cluster.

Recently, the gross spectrum from clusters of galaxies was shown to be
inconsistent with simple models of cooling flows 
\citep[e.g.\ ][]{peterson1}. The confusion over what this result means clearly
precludes the use of any simple theoretical template in the construction
of a model of the intracluster medium.   Our goal in this paper is
therefore to provide a flexible method that is capable of reproducing
all of the complexities we have mentioned above.  This method must
necessarily utilize both the spectral and spatial information in a
dataset and should still be effective with datasets that have relatively
low numbers of X-ray counts.

In the literature is a large number of diverse techniques for analyzing
X-ray data from galaxy clusters.  The methods range from fitting
isothermal spectra to square grids of photons across the detector plane
\citep[e.g.\ ][]{markevitch} to imaging deprojection techniques through
``onion-peeling'' methods \citep[e.g.\ ][]{fabian_depro}.  Fitting
adaptively-binned data to isothermal models has also proved useful
\citep{sanders} and the onion-peeling technique has been extended to
spectroscopic deprojection \citep{arabadjis,arnaud,kaastra,andersson}.  Other
methods include image smoothing, gross spectral fitting, and inversion of
spectral hardness ratios \citep{sanderson}. 

We break with these previous methods in two important ways. First, we
use the spectrum and image of a source on an equal footing.  Traditional
analyses extract a spectrum from a piece of the detector plane and then
use the integrated number of photons as a proxy for the true source
image. However, this process is only approximately correct in the
presence of broad point-spread functions that are energy-dependent. 
Even the concept that there is an ``image'' and a ``spectrum'' for a
spatially-resolved source is misleading, since in reality there is a
spatially-varying emissivity at a given energy. We propose to use a
Monte Carlo calculation to forward-fold a complex model to predict
detector positions and CCD energies for each photon.  Secondly, we
construct a model that is flexible enough to allow for extremely complex
gas temperature and density distributions in clusters.  For example, a
common approach is to assume a single temperature for a given
set of photons and fit for its value. In contrast, our procedure is to
allow for a multi-temperature distribution, and then allow the method to
sharpen the temperature distribution if it is close to a single
temperature. This approach is motivated by the observations outlined in
the first paragraph that clusters are complex. We achieve this
complexity by modeling the emissivity from clusters as using 
conglomerations of emissivity particles, smoothed for ease of
interpretation.  

The construction of underlying model distributions from  linear
combinations of simple basis functions is well-founded in modern
Bayesian image processing: indeed, the ``massive inference'' method of
\citet{massinf} has been found to be a powerful technique for
the reconstruction of one dimensional spectra and two dimensional
images. This work may be seen as a practical application of the basic
premises of massive inference to the problem of astronomical X-ray data
analysis, with some of the refinements compromised in the name of
pragmatism. Skilling et al. coined the term ``atom'' for the indivisible
unit of image flux; to avoid confusion with the atomic physics
terminology of X-ray spectroscopy we translate atom into particle and
draw an analogy with numerical simulations of galaxy clusters. Where in
the first instance our goal is to model X-ray data, it will become clear
that this ``smoothed particle inference'' (SPI) methodology offers great
opportunities for further analysis and physical interpretation.  This work
also borrows heavily from some independent pioneering work by
\citet{jernigan}.  There a Monte Carlo approach was used to invert X-ray data
by interating a non-parametric model, while matching the simulation with the
X-ray events.

In this paper, we describe the SPI method in considerable detail,
highlighting the two novel aspects introduced above. We then  apply the
method on realistic simulated  datasets for both EPIC and RGS, and show
that it works. We then demonstrate some of the benefits of the
methodology with regard to extracting and displaying the information in
the data.

In two companion papers, we apply the method to XMM-Newton Reflection
Grating Spectrometer (RGS) observations of the Perseus and Virgo
clusters \citep{peterson4} and XMM-Newton European Photon Imaging
Camera (EPIC) observations of Abell 1689 and RXJ0658-55
\citep{andersson2}.


\section{Methodology}

This section is organized as follows.  We first describe our choice of
model, a set of spatially Gaussian X-ray plasma emitters of an assigned
spectral type.   Then, we discuss X-ray Monte Carlo techniques to
propagate this model through the instrument response.  We discuss the
ensuing two-sample likelihood statistic used to assess the statistical
significance of the model.  Finally, we describe the use of Markov Chain
techniques to iterate the model and explore the parameter space.


\subsection{Smoothed Particles in a Luminosity Basis}

We have argued in the previous section that a ``non-parametric'' method,
or rather, a method with thousands of parameters, is required to
describe the current X-ray data.  In particular, a more unified modeling
approach, embracing all the complex aspects of clusters that have
recently been discovered (cold fronts, temperature radial gradients,
local non-isothermality, lack of circular symmetry in the temperature
distribution) is now required. 

Consider a ``blob'', or ``smoothed particle'', of plasma, described by a
spatial position, a Gaussian width, a single temperature, a set of
elemental abundances, and an overall luminosity.  A model of a cluster
of galaxies may be  constructed from  a set of such particles. There are a
number of advantages to doing this.  The model has enough freedom to
reproduce all of the salient features of cluster gas distributions, and
can form complex density and temperature structures.  Particles with
different temperatures can occupy the same position and therefore
multi-phase temperature distributions can be constructed.  There is also
no global spherical symmetry that is imposed on the particle configuration,
although they are themselves spherically symmetric.  Another powerful
aspect of using smoothed particles is that they can be mapped from one
space to another in a straightforward way: for example  a collection of
smoothed particles representing X-ray luminosity may be manipulated in
order to derive the corresponding SZ decrement map.  This approach is
somewhat similar to the methods used in smoothed particle hydrodynamics
(SPH), except here the method decide the properties of the smoothed
particle, rather than a set of hydrodynamical equations.  For this
reason, we refer to the method as ``smoothed particle inference''.

Assuming an X-ray source is optically-thin, we can write the
differential  luminosity per volume due to a set of smoothed Gaussian
particles as

\begin{equation}
\frac{d^2L}{dV dT} = \sum_i L_i \delta\left(T-T_i\right) 
\frac{e^{-\frac{\left|{{\bf
          x}-{\bf x_i}}\right|^2}{2 {\left(D_A \sigma_i \right)}^2}}}{\left(2 \pi {\left(D_A
  \sigma_i \right)}^2 \right)^{3/2}}
\end{equation}

\noindent where $L_i$ is the luminosity per particle in ergs per second, $T_i$ is the
temperature of the $i$th particle in keV, $D_A$ is the angular diameter
distance in $\mbox{cm}$, $\sigma_i$ is the angular size in radians, ${\bf x_i}$ is the location of
the particle in three dimensions expressed in $\mbox{cm}$.  We achieve the differential
luminosity per solid angle by integrating along the line of sight,
$dz$, 

\begin{equation}
\frac{d^2L}{d\Omega dT} = D_A^2 \int_{-\infty}^{\infty} \frac{d^2L}{dV dT} dz.
\end{equation}

\noindent This can be rewritten as

\begin{equation}
\frac{d^2L}{d\Omega dT} = \sum_i L_i \delta\left(T-T_i\right) 
\frac{e^{-\frac{\left|{{\bf
          x}-{\bf x_i}}\right|^2}{2 {\left(D_A \sigma_i \right)}^2}}}{\left(2 \pi {\left(D_A
  \sigma_i \right)}^2 \right)}
\end{equation}

\noindent where ${\bf x_i}$ is now the two dimensional spatial position
of each smoothed particle.  The differential flux per unit solid angle
per unit energy,  $\frac{d^2F}{dEd\Omega}$, in units of photons per second per
$\mbox{cm}^2$ which is the actual
observable, is then obtained by first calculating the differential flux
per unit energy, $\frac{dF}{dE}$, 

\begin{eqnarray}
\frac{dF}{dE} & = & \frac{1}{4 \pi D_L^2} \int_0^{\infty} dT \frac{d\alpha}{dE} \frac{dEM}{dT}
\nonumber \\
&  =  & \frac{1}{4 \pi D_L^2} \int_0^{\infty} dT \frac{d\alpha}{dE} \frac{dL}{dT} \frac{1}{\Lambda} \nonumber
\\
\end{eqnarray}

\noindent where $D_L$ is the luminosity distance in $\mbox{cm}$, $\Lambda$ is the
cooling function in units of $\mbox{ergs} \mbox{cm}^3 s^{-1}$, $\frac{dEM}{dT}$ is the differential emission measure,
and $\frac{d\alpha}{dE}$ is the line (or continuum power) which is
predicted uniquely given a set of elemental abundances and a
temperature in units of $cm^3 s^{-1} keV^{-1}$.  Note that the cooling function is the energy averaged
integral of $\frac{d\alpha}{dE}$.   The differential flux per solid
angle per energy is then obtained by combining equation 3 and equation
4, 

\begin{eqnarray}
\frac{d^2F({\bf \theta},E)}{dE d\Omega} = \sum_i \frac{L_i}{4 \pi D_L^2
  \Lambda(T_i, \{A_i \})}  \times \nonumber \\
\frac{d\alpha}{dE} (E | T_i, z_i,
\{ A_i \} )~\frac{e^{-\frac{\left({{\bf
          \theta}-{\bf \theta_i}}\right)^2}{2 \sigma^2}}}{\left( 2 \pi
    \sigma^2 \right)}.
\end{eqnarray}

\noindent where ${\bf \theta}$ are angular coordinates on the sky,
$\sigma$ is now in angular units, $D_L$ is the luminosity distance in $\mbox{cm}$,
$\Lambda(T_i, \{A_i \})$ is the temperature and metallicity-dependent
cooling function, and $\frac{d\alpha}{dE}$ is the energy-dependent line
(or continuum) power function, which describes the probability of a
photon having a given energy~$E$ as a function of temperature.  

We have purposely chosen to use luminosity as the basis for our smoothed
particles, rather than, say, gas density.  If we had parameterized in
terms of the gas density then predicting the luminosity would require,
essentially, a summation over all particles to reconstruct the density
field and then the squaring of the result.  In this case, the complete
luminosity field would have to be reconstructed following the alteration
of a single particle. Later, we will demonstrate that using a luminosity
basis will simplify the calculation considerably but not quantitatively change
the results.  We are also
anticipating that in using this method in combination with SZ
observations, it will still be preferable to use the luminosity as the
basis rather than a Compton-y parameter basis since the calculation of
the X-ray instrument response is likely to be much more involved.


\subsection{X-ray Monte Carlo Methods}

We convert the luminosity at each energy and spatial position into a
quantitative prediction for probability of the numbers of photons at a
given detector position and energy via a Monte Carlo calculation.  More
explicitly, our goal is the calculation of the probability of
detection,~$D$, given the instrument response function,~$R$, and the
photon source model,~$S$.  This is expressed as,

\begin{equation}
D ({\bf u},p) = \int dE~d\Omega ~ R ({\bf u},p~|~ E,{\bf \Omega}) ~
\frac{d^2F}{dE d\Omega} (E,{\bf \Omega})
\end{equation}

\noindent where ${\bf u}$ are the detector position vector,  $p$ is the
measured detector energy, $\Omega$ is the sky position vector, and $E$
is the intrinsic energy of the photon.  This is a three-dimensional
integral: it is most efficiently calculated using a Monte Carlo
approach.  In \citet{peterson3}, we describe the calculation of this
integral for the Reflection Grating Spectrometers (RGS) on XMM-Newton. 
We have also written a companion Monte Carlo for the European Photon
Imaging Cameras (EPIC) on XMM-Newton \citep{andersson2}.  We do not
describe these calculations in detail here, but instead briefly outline
their use in the context of smoothed particle simulation.

In the Monte Carlo calculation, photons are simulated sequentially via
conditional probability functions.  Mirror, grating, and detector
responses are included in the calculation.  For the RGS simulation,
photons are manipulated using a set of dispersion and cross-dispersion
response functions, CCD pulse-height redistribution functions, and
exposure maps.  For the EPIC Monte Carlo, photons are constructed using
the point-spread function, vignetting, CCD pulse-height redistribution
functions, and the exposure map.  In general, the functions are both
energy and off-axis angle dependent and the calculation of these
functions involves a sequence of convolutions.  The response functions
are calculated on coarse grids, and then interpolated using a Monte
Carlo method \citep{peterson3}.  Photons are removed in proportion to the
relative response probability, in
order to properly calculate the effective area, which is also off-axis
angle and energy dependent.  

For the purpose of simulating the X-ray emission from a set of SPI
particles, a set of simulated photons is assigned to a particular particle. 
The particle parameters are the temperature ($T$), set of abundances,
position ($\phi$, $\psi$), Gaussian width ($\sigma$), and possibly the
redshift.  For each photon, the wavelength is first chosen from the
probability distribution described by the emissivity function
corresponding to the temperature and set of elemental abundances of that
particle.  Its position is then perturbed from the nominal center of that
particle ($\phi$, $\psi$) by a random Gaussian step having width equal to
$\sigma$. 

The photon flux per particle, $F_i$, is calculated by computing,

\begin{equation}
F_i= \frac{N}{A_{\rm eff} \tau}
\end{equation}

\noindent where $A_{\rm eff}$ is the peak effective area in units of 
$\mbox{cm}^2$, $\tau$ is the
exposure time in units of $s$, and $N$ is the number of photons simulated before
removing any photons in the Monte Carlo calculation.  The luminosity,
$L_i$ per particle can be calculated by computing,

\begin{equation}
L_i = F_i \frac{\Lambda_i}{P} 4 \pi D_L^2
\end{equation}

\noindent where

\begin{equation}
P= \int dE \frac{d\epsilon}{dE}
\end{equation}

\noindent and $\frac{d\epsilon}{dE}$ is the emissivity per energy
interval in units of $\mbox{cm}^2 \mbox{s}^{-1} \mbox{ergs}^{-1}$. 


\subsection{The Likelihood Function}
\label{sect:likelihood}

We quantify the goodness of fit of a given set of SPI particles via the 
likelihood function of their parameters. In general, the likelihood is a
function dependent on both predicted and actual data, whose form
embodies the analyst's understanding of the measurement errors, and is
understood as the probability of getting the data given the parameters
of the model. Ordinarily, the predicted data are calculated to arbitrary
precision within the context of the model under investigation: this is
not the case here. The Monte Carlo simulation procedure outlined above,
while needed to deal  correctly with the energy-dependent and spatially
varying telescope response, provides predicted data that are themselves
uncertain, being the outcome of a Poisson process. We can view the
output of the telescope, and the output of the simulation, as two
independent data streams emanating from the same underlying physical
process (emission from the cluster followed by detection at the
telescope). The likelihood function then quantifies the misfit between
the two datasets. 

Note that given infinite simulation time, predicted data can be computed
with infinite precision within the context of the limitations of the
instrument model; clearly any likelihood statistic used should revert to
the more familiar form in this limit. Moreover, we note that all
likelihood functions are necessarily approximations to an unknown
function (that indeed, should itself form part of the data model). We
should be satisfied with a likelihood that captures the essence of the
misfit, and satisfies any limits such as that above.  Statistics for the two Poisson sample situation have not been
well-developed in the statistical literature; possibly it has been
deemed preferable to find alternative ways of computing predicted data
exactly rather than press on with effectively two sources of measurement
error. Where attempts have been made, they have made use of Gaussian
approximations valid only in the limit of large numbers.  In the
appendix we give a discussion of a number of routes to a likelihood
function applicable to the low counting statistics of X-ray datasets,
and show how they recover the correct form in the limit of infinite
simulation time. We give below  the form of the likelihood~$L_2$ that 
we use in practice, and note that in its normalized form ($L_{2N}$)  that
$-2 \log L_{2N}$ is distributed as $\chi^2$ with the number of degrees of
freedom equal to the number of bins. while not required by the
analysis, we find that this provides a valuable check on the pure
goodness of fit of the models, giving confidence in the results. 

In general, we find that the results are relatively insensitive to the
exact choice of likelihood statistic and the differences are largest
when the data is extremely sparse and there are few photons: in
subsection~\ref{sect:adbin} below we outline our approach for avoiding this
eventuality.  We experimented with the likelihood functions in the next
section and found the reconstruction to produce similar results.


\subsubsection{Two Sample Likelihood Statistic}

The basic datum is a number of photons in a three dimensional bin $(u_1,
u_2, p)$. The bin size may be set by physical limits such as the extent
of the CCD  pixel, or be imposed given some understanding of, for
example, the spectral resolution of the instrument. The simulated
photons are then binned on the same grid. Assume that we have $n$
photons in the observed dataset, and have simulated $m$ photons from our
model: in the $i^{\rm th}$ bin we have  $E_i$ simulated photons, and
$O_i$ observed photons.  Since the photons falling into a set of bins
will follow a multinomial distribution, we obtain for the data photons
the likelihood function

\begin{equation}
L_{O} = n ! \prod_i \frac{{\left( \lambda_i \right)} ^ {O_i}}{O_i !}
\end{equation}

\noindent where $\lambda_i$ is the probability of getting a photon in that
bin.  The multinomial distribution is the Poisson distribution but with total number of photons fixed.  Treating the simulated photons in the same way, we get

\begin{equation}
L_s = m !  \prod_i \frac{{{\left( \lambda_i \right)}}^{E_i} }{E_i!}
\end{equation}

\noindent In doing so we have suggested that the simulated photons be
considered to be from a separate, independent, experiment with its own
likelihood.  The joint likelihood is then just the product of the two
experiments' likelihoods. In the appendix we discuss two routes to
dealing with the value of the parameters $\lambda_i$: one can
see that a reasonable guess is to estimate $\lambda_i$ from the data and
set $\lambda_i = (O_i + E_i)/(m + n)$: 

\begin{equation}
L_2 = n ! m !  \prod_i \frac{{{\left( \frac{O_i+E_i}{m+n} \right)}}^{O_i} 
{{\left( \frac{O_i+E_i}{m+n} \right)}}^{E_i} }{O_i ! E_i!}
\end{equation}

\noindent We use this form when comparing exploring the parameter space
as outlined in the next section. It is possible to re-normalize these
likelihood functions, such that they will approximately follow the
$\chi^2$ distribution~\citep{Cowan}. This is done by dividing by the same
expression with the probability estimates replaced by the actual number
of counts.  The result for the likelihood given above is written as

\begin{equation}
L_{2N} = \prod_i \frac{\frac{{{\left( \frac{O_i+E_i}{m+n} \right)}}^{O_i} 
{{\left( \frac{O_i+E_i}{m+n} \right)}}^{E_i} }{O_i ! E_i!}}{\frac{{{\left( \frac{O_i}{n} \right)}}^{O_i} 
{{\left( \frac{E_i}{m} \right)}}^{E_i} }{O_i ! E_i!}}
= \prod_i \frac{{{\left( \frac{O_i+E_i}{m+n} \right)}}^{O_i+E_i} }
{{{\left( \frac{O_i}{n} \right)}}^{O_i} 
{{\left( \frac{E_i}{m} \right)}}^{E_i} }
\end{equation}

\noindent By construction, $-2 \log{L_{2N}}$ is approximately distributed
as $\chi^2$ with the number of degrees of freedom equal to the number of
bins.  We use $L_{2N}$ for checking that a reasonable fit is being
obtained by the method (see also section~\ref{sect:evidence}), which will be
near the number of degrees of freedom for a good fit.  In the
appendix we demonstrate how this statistic approaches the two sample
$\chi^2$ statistic that we used in the large $m$ and $n$ limit.  $L_2$ is used
for the parameter iteration, since it analgous to the one-sample likelihood.


\subsubsection{Three-Dimensional Adaptive Binning}
\label{sect:adbin}
The statistics that we have discussed in the previous section require
the events to be binned on a arbitrary three-dimensional grid.  For a
non-dispersive spectrometer, the natural minimal grid for this would be
to design bins to reasonably sample the point-spread function for the
two spatial dimensions and the energy resolution kernel for the spectral
dimension.  For a dispersive spectrometer, the angular dispersion
resolution kernel would define the bins in the dispersion direction.  In
either case, however, it is found that, for virtually all observations
of clusters of galaxies, the numbers of minimal bins are very large
indeed, and that the photons fill the three dimensional grids extremely
sparsely.   In fact, most bins typically contain either 0 or 1 photons. 
Simple grouping of these minimal bins will result in severe loss of
spectral and/or spatial resolution. Therefore, a more efficient binning
approach deserves some consideration.  Although the statistics discussed
in the previous section are capable of handling low numbers of photon
counts, we are pushed toward fuller bins to reduce  computation time. 
Additionally, we show below that the fluctuation induced on the
statistic by the Monte Carlo calculation depends on the number of bins,
such that a smaller number of bins is preferred.

For these reasons, we have developed an algorithm to adaptively bin the
data in the three dimensions similar to other adaptive binning methods. 
To do this, we first consider the entire three-dimensional data space. 
We create the binning  grid by bisecting the data space in one dimension
at a time.  The dimension to bisect is chosen from the relative length
of the bin in each dimension.  The longer dimension is always bisected. 
If the relative sizes are the same the  dimension is selected randomly. 
Each bin is bisected for as long as any bin has more than $N$ counts;
most of the bins will contain on average $N/2$ photon counts by the
time the binning has been completed.  Typically, $N$ is chosen to be 10
or 20, but this varies between applications.  For instance in a dataset
with extremely low flux the bins can be made with fewer photons.  There
is also the possibility of weighting the bin size in any dimension with
some factor $n$.  For example, if spectral accuracy is preferred over
spatial accuracy  the bins in the spectral dimension can be chosen to be
on average $n$ times smaller than the spatial bins. After the bins have
been determined for the data photons, they remain fixed.  We then use a
binary tree algorithm to rapidly place new simulated photons into the
appropriate bin.  The two-sample statistic given in the previous section
can then be calculated as a product over all the bins.


\subsection{Parameter Iteration}

After model generation and likelihood calculation, the third major
component of the method is the sequence of model parameter iteration. 
We have found several complications to the usual methods of parameter
iteration that are particularly problematic with the high-dimensional
parameter model that we are considering in this paper.  In the 
following subsections we discuss several aspects of this part of the
analysis chain.


\subsubsection{Markov Chain Monte Carlo Exploration}

We explore the parameter space of the smoothed particle model with the
Markov Chain Monte Carlo (MCMC) method, the use of which has recently
become widespread in the field of Bayesian data analysis 
\citep[see ][ for an excellent introduction]{mcmc}.  
MCMC is a technique for sampling
the probability distribution of a model's parameters; in this case, the
dimensionality of the parameter space is extremely high, prohibiting
brute force grid calculations and condemning gradient-based optimization
to failure by being trapped in local likelihood maxima. MCMC is a very
efficient and powerful way of exploring such high-dimensional spaces,
returning a list of sample cluster models all of whom are consistent
with the data, and whose number density in parameter space follows the
probability density.

Possession of the entire parameter probability distribution, in the form
of a list of samples, opens up possibilities not available to other
methods. The distributions of particle parameters, for example, can be used
to estimate the average distributions of cluster quantities (such as
temperatures and densities).  These averages come with well-defined
uncertainties trivially derived from the samples: one may probe the
range of distributions that are consistent with the data. MCMC is now
the standard tool for cosmological data analysis
\citep[see e.g.\ ][]{lewis+bridle,tegmark}.  
MCMC has been applied to simple cluster modeling using SZ and 
lensing data
\citep{marshall2}, and X-ray and SZ data \citep{bonamente}. 
To date, the use of MCMC in astrophysics has
been restricted to relatively modest numbers of parameters. However, it
is when using large numbers of parameters to model complex datasets that
the full power of the technique is realised.

As usual, the target probability distribution is the posterior
density~$P$, given by Bayes' theorem as 

\begin{equation}
P(\{\theta\} |  D) \propto L(D|\{\theta\}) \pi( \{ \theta \}).
\end{equation}

\noindent Here, $L$ is the likelihood function introduced above, with
the notation emphasizing its dependence on the observed data $D$ given
the parameter set $\{ \theta \}$. $\pi$ is the prior probability
distribution of the model parameters: in this work we use uninformative
priors on all parameters, restricting ourselves to uniform
distributions over broad finite ranges. 

We give here the briefest of introductions to the Metropolis-Hastings
algorithm \citep{ST/Met++53,ST/Has70} that we use to sample the posterior density, and
then provide details of the implementation in the following
subsections. Following the evaluation of the likelihood associated with
an initial point in parameter space, $P_1 \equiv \left( \theta_1,
\theta_2, \ldots, \theta_n \right)$,  a new point, $P_2$ is drawn from
a proposal distribution.  This proposal distribution is required to be
symmetric about the current position to ensure successful sampling of
the target distribution. If the likelihood of the new point is higher
than the likelihood of $P_1$ then the new point is accepted. 
Otherwise, the point is accepted only if a uniform random number
between 0 and 1 is less than the ratio of likelihoods,
$\frac{L(P_1)}{L(P_2)}$.  This is known as the Metropolis-Hastings
acceptance criterion.   If the new point is not accepted, then the old
point is repeated in the chain of parameters. While this basic
algorithm is very simple, the  efficiency of the  exploration of the
parameter space depends strongly on the  form of the proposal
distribution, which we discuss in the following two subsections. 


\subsubsection{Adaptive Proposal Distribution}
\label{sect:proposal}

As we mentioned in the previous section, the method of picking the next
point in the Markov chain can either be by taking a step from some fixed
distribution or it can be modified dynamically.  In dealing with the
large number of parameters in the smoothed particle fits, we have found
that an adaptive proposal distribution is necessary.  Initially, the
proposal distribution is chosen to be a Gaussian of fixed width,
$\sigma$.  After the Markov chain has taken a few steps, we calculate
the standard deviation, $\sigma^{\prime}$, of the new estimate of the
target posterior distribution.  We then use $\sigma^{\prime}$ as an
estimate for the new step in the proposal distribution.  Large
excursions in parameter space, however, can sometimes bias this standard
deviation estimation. Therefore, we have constructed a different
standard deviation estimator with some built-in ``memory loss'', such
that earlier steps are weighted less strongly. The proposal step width,
$\sigma^{\prime \prime}$ is calculated according to

\begin{equation}
  \sigma^{\prime \prime} = \left( \frac{ \sum_i (1+\epsilon)^i p_i^2  }{ \sum_i
    (1+\epsilon)^i} - \frac{ \left( \sum_i (1+\epsilon)^i p_i \right)^2 } {
    \left( \sum_i (1+\epsilon)^i \right)^2 } \right)^{\frac{1}{2}}
\end{equation}

\noindent where $p_i$ is the parameter value at the $i^{\rm th}$
iteration and the sum is over all the iterations.  $\epsilon$ is a
constant taken to be $0.01$ that causes the earlier iterations to be
de-weighted.  If $\epsilon$ were zero, the formula would result to the
normal standard deviation formula; an infinite value for epsilon 
implies the normal non-adaptive Gaussian proposal distribution.  With
finite $\epsilon$, the individual steps no longer exactly satisfy
detailed balance, as the proposal distribution is slightly different
between the current and trial sample positions.  If epsilon is set
sufficiently small (although not so small as to restrict the movement of
the chain) then a large number of past sample positions are used to
determine the proposal distribution widths, and consequently the
differences between any successive proposal distributions are very
small.  The parameter update method then asymptotically becomes a true Markov chain.  In practice these differences are completely negligible compared
to the large deviations from unity of the likelihood ratios.  Quantitatively,
we found that this holds whenever $2 \Delta \log{L} \gg \frac{\Delta
  \sigma^{\prime \prime}}{\sigma^{\prime \prime}} \approx \epsilon$.  In
practice $2 \Delta \log{L}$ is much larger than one and we take $\epsilon$ to
be 0.01.

The proposal step is then taken to be $2.38 \sigma^{\prime \prime}
d^{-\frac{1}{2}}$, where $d$ is the number of dimensions.  This is the
optimal step size for uncorrelated Gaussian distributions, such that the
acceptance rate times the RMS step size is maximized for this value.  In the large $d$ limit, a rejection rate around
77$\%$ is expected \citep{roberts}. A successful chain will, after the burn-in period,
explore the posterior distribution with this rejection rate,  and the
proposal distribution in any given parameter direction will asymptote to
the  Gaussian approximation to the marginalized posterior.


\subsubsection{Global and local parameters iteration sequence}

In a typical SPI emission model there are two kinds of parameters. The
first kind are local parameters, describing the individual properties of
the particles.  These are the temperature, position, size, and elemental
abundances for each particle.  The second kind are global
parameters for describing properties of the entire data set.   These
parameters are the background parameters and the absorption column density. The
order in which these parameters are varied was found to have a strong influence
on the speed of the chain convergence: the variation sequence has been
under considerable study in the development of this method. 

Firstly, it was found that the local parameters for a particular particle
should be varied separately while holding the parameters of the other
particles fixed and the global parameters fixed.  We believe that this is a
valid technique since any particle can be interchanged with any other particle
and therefore we can exploit this symmetry by not considering the full
degeneracy of all of the particles parameters.  We do, however, vary the 
set of particle parameters -- position, temperature, elemental abundances,
and Gaussian size-- simultaneously since these parameters may be highly
correlated.  Varying each set of particle parameters separately  speeds up
the calculation by several orders of magnitude, as only the photons
associated with that particle need to be resimulated.  

Secondly, we found that after changing the global parameters the particle
parameters needed to be varied for at least one iteration in order for
the system to re-equilibrate and remain in the desired high probability
region. This is necessary since there may be some correlations between
the positions of the particles, for example, and global background
parameters.  Therefore, the particle parameters must be allowed to move
somewhat before rejecting a new set of global parameters.  Without this
consideration, we found some global parameters became pinned to
sub-optimal values. 

Thirdly, we found it necessary to regenerate a completely new set of
photons for a particular set of parameters frequently.  This is
necessary due to the Monte Carlo nature of the calculation: there is
statistical noise in the likelihood value itself.   Although we believe
the two sample likelihood properly deals with this statistical noise in
modifying the likelihood surface, it does not deal with the parameters
drifting to a particular non-optimal value only because it received a
favorable set of photons.  Frequent regeneration of the photons simply
cancels this effect and restores the mobility of the Markov chain.

These three considerations lead us to the following sequence of
parameter iteration.  Consider that we have $n$ particles with local
parameters, $l^i_j$, for the $i$th particle and the $j$th particle parameter. 
Assume also that we have a set of global parameters $g_k$ for the $k$th parameter.  The
following sequence of parameter variation is then used for each iteration:

\begin{enumerate}

\item{Vary values of $l^i_j$ to new values $l^{\prime i}_j$, holding the
  parameters of all other particles fixed, and compute the likelihood.  Take the
  new values if the statistic conforms to the acceptance criterion.  Repeat
  this step for each of the particles.}

\item{Regenerate all photons for $g_k, l^{\prime i}_j$ and compute the likelihood.}

\item{Vary the values of $g_k$ to a new point $g_k^{\prime}$.  Hold the
values of $l^i_j$ fixed at the previous values.}

\item{Vary the values of $l^i_j$ to new values $l^{\prime \prime i}_j$,
  holding the parameters of all other particles fixed, and compute the
  likelihood.  Take the new values,
  if the statistic conforms to the acceptance criterion.  Repeat this
  step for each of the particles.}

\item{Regenerate all of the photons $g_k^{\prime}$ and possibly the new values
  of $l^i_j$.  Compare the value of the likelihood at this point
  with that obtained in the second step.  Take the new value of $g^{\prime}_k$
  and values of $l^{\prime \prime i}_j$,
  if it conforms to the acceptance criterion.  Otherwise, return to the previous
  values of both $g_k$ and $l^{\prime i}_j$ at the second step.}

\end{enumerate}

The absence of a Metropolis comparison in step 3 indicates a departure
from the standard Markov chain: in fact, this step can be thought of as
the splitting off of a new chain which explores the posterior
conditional on the value of $g_k^{\prime}$. After some exploration of
the local parameter space, this chain is then compared with its parent
(left behind at the known sample position $g_k$) and the fittest (in the
Metroplis-Hasting sense) survives in step 4. Strictly speaking this
process does not conserve detailed balance: however, we have checked
that the departure of this algorithm from the proper format does not
affect the convergence on the target posterior distribution, at least in
simple cases as we demonstrate in Appendix 2.


\subsubsection{Increase of the Over-simulate factor and Statistical Noise}

\begin{figure*}[!htb]
   \begin{center}
     \begin{tabular}{cc}
       \includegraphics[width=2.5in,angle=-90]{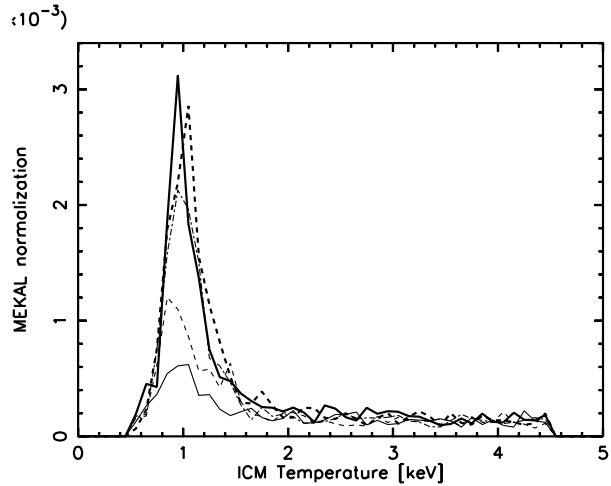} \\ 
     \end{tabular}
   \end{center}  
   \caption{Shown is the resolution of the 1 keV temperature component from the RGS run for different values of the over-simulate factor. The lines respresent over-simulate 1 (solid-thin), 3 (dashed-thin), 10 (dot-dashed), 30 (solid-thick) and 100 (solid-dashed). The necessity of a high over-simulation ($\sim 10$) for these data is clear. \label{fig:oversim}}
 \end{figure*}

Since we have used a Monte Carlo calculation to calculate the likelihood
statistic for the highly-complex smoothed-particle model, there is
unavoidable  statistical noise in the likelihood function.  The same set
of parameters will produce a slightly different value of the likelihood
statistic every time they are simulated.  This Monte Carlo noise can be
reduced by increasing the ``over-simulate factor,'' the ratio of the
simulated photons compared to the number of photons in the data set.  In
practice, a value of the over-simulate factor of 10 or so can be easily
accommodated, but much larger values become tedious when working  on a
single workstation.  The statistical fluctuation in the two sample
normalized statistic scales empirically as

\begin{equation}
\sigma_{L_{2N}} \sim \sqrt{\frac{\mbox{number~of~bins}}{\mbox{over-simulate~factor}}}
\end{equation}

\noindent such that the fluctuation can be reduced by both increasing
the simulation factor and reducing the number of bins.  Normally, this
ratio is significantly larger than one, and therefore this is likely to
significantly broaden the posterior distribution of any
parameters.  However, this merely makes the method conservative: the
posterior distributions we derive contain the smaller true
distribution.  This is illustrated in Figure~\ref{fig:oversim}.


\subsubsection{Particle Number and the Evidence}
\label{sect:evidence}

Initially, we experimented with methods that may change the number of
particles dynamically, during the evolution of the  Markov chain.  However,
we found that any advantage that this method may have had, in reducing
the number of smoothed particles and their associated  parameters, was
outweighed by the complications added to the fitting procedure. In
particular, it was difficult to determine what to do with the parameters
of the particles that were newly created or recently destroyed when the
particle number changed.  The use of adaptive steps within this scheme was
also problematic.  Therefore, we fix the number of particles and do not
vary this during the parameter iteration, instead opting for a more
approximate approach to determining a suitable number of particles.   

We calculate (albeit approximately) the ``evidence'' of the smoothed
particle model as a function of particle number.   The evidence is the
integral of the likelihood function $L$ times the prior distribution, $f(p_i)$ over
all parameters($p_i$): 

\begin{equation}
E = \int \prod_i dp_i L f(p_i)
\end{equation}

The evidence works by penalizing overly complex models (such as those with too many
particles) wwith a lower evidence
\citep[see e.g.\ ][ for introductions to the use of the evidence]{mckay,oruanaidh,marshall2}.
We estimate the evidence by the method of thermodynamic
integration \citep{oruanaidh}, raising the likelihood to some power
during the initial stages of the MCMC process and increasing this power
steadily from 0 to 1.  In our studies, the evidence increased
dramatically with increasing particle number for less than $\sim100$
smoothed particles but was then relatively flat for larger particle
numbers. This number of course varies from dataset to dataset: we use
the smallest number of particles possible (indicated by the turnover in
evidence, where the improvement in goodness-of-fit starts to become
insignificant) in the interests of both the economy of hypotheses and 
also of saving computation time.

The method requires one to choose both the over-simulate factor and the
number of smoothed particles. The former is chosen to be as large as
possible without the method taking too long, and the latter is chosen to
be as large as possible given the criteria of the previous paragraph. 
Having finite values for the over-simulate factor limits the precision
of the method, but this was found to be unimportant for many
applications. Limiting the number of particles via consideration of the
evidence prevents fitting of the noise with an overly complex model.


\section{Model Interpretation Methods}

There is an impressive variety of techniques that can be applied to the
results of the smoothed particle inference.  The output of the MCMC
process consists of a set of  parameters, both global and local, for the
given number of smoothed particles, for each iteration.  Thus, the usual
marginalized  posterior probability distributions can be generated by
histogramming the history of parameter values (see
Fig.~\ref{fig:histopars}).  The interpretation of the smoothed particle
parameters is not always straightforward, but a wealth of information
resides in the MCMC samples. SPI is essentially a method for translating
noisy X-ray data into uncertain emission components: the Markov chain
process retains the correlations between components such that the
manipulation of the samples can reveal much about the gross and detailed
properties of the emitter.  We discuss just some of the possible
inference techniques below. As always, caution has to be used in only
using samples taken after the chain has reached the stationary
distribution.  

To demonstrate these inference techniques we have simulated a cluster 
model with two spatially separated isothermal components to which we
apply  our method. 
This choice of simplistic simulation reflects both our desire to
demonstrate the algorithm on a relatively easily-interpreted problem,
but also the difficulty of generating realistically complex mock
datasets.
To make a simple demonstration of the features of the method we have 
chosen to simulate a double cluster with two isothermal components. 
This will test the ability of the method to estimate single temperatures as 
well as to separate different components in a dataset, both spatially 
and spectrally. The method is applied to real datasets in two companion 
papers, \citet{peterson4} and \citet{andersson2}.

In this work we perform separate inferences for the XMM-Newton EPIC 
and XMM-Newton RGS detectors. 
In the EPIC run we expect to be able to resolve the spatial structure of
the clusters in detail due to the high spatial resolution (PSF FWHM
$\sim 6''$)  of the instrument datasets.  For the RGS run the spatial
resolution will be low in the dispersion direction of the spectrometer;
however we expect to resolve spectral features in finer detail.


\subsection{A simulated double-cluster model}

The spatial model of the simulated cluster consists of two beta-models,  

\begin{equation}
S(r) = S_0 \left( 1 + \left( \frac{r}{r_c} \right)^2 \right)^{-3 \beta+0.5}
\end{equation}

\noindent where $S(r)$ is the source surface brightness at angular radius $r$ 
and $r_c$ is the angular core radius. The beta models are separated diagonally
by $1'$.  Each of the two clusters is modeled  spectrally with a MEKAL
\citep{mekal1,mekal2,mekal3,mekal4} model for the thermal plasma with 
Wisconsin \citep{wabs} photo-electric absorption for the Galactic column. 
The cluster plasma temperatures
were chosen separately for the  EPIC and RGS simulations due to the
different energy sensitivity  of the two instruments.  In the RGS run
the two beta models have isothermal temperatures of \TkeV{1} and \TkeV{3}
respectively whereas in the EPIC run these temperatures are at \TkeV{3}
 and \TkeV{6}. 

Metal abundances are set at 0.5 Solar, the redshift is fixed  at
$z=0.2$ and both clusters are  absorbed by a corresponding Hydrogen
column of  $4 \times 10^{20}~\mathrm{cm^{-2}}$. The beta model
parameters for the clusters were set to  $\beta = 0.7$, $r_c = 40''$ and
$\beta = 0.5$, $r_c = 15''$ for the lower left and the upper right
cluster respectively (see Fig.~\ref{fig:lummed}).

The numbers of simulated photons for the cluster model were 155000 for
the EPIC simulation and 55000 for the RGS simulation, corresponding to 
those of a well exposed, bright cluster at $z = 0.2$. 

\begin{figure*}[!htb]
  \begin{center}
    \begin{tabular}{cc}
      \includegraphics[width=2.5in,angle=-90]{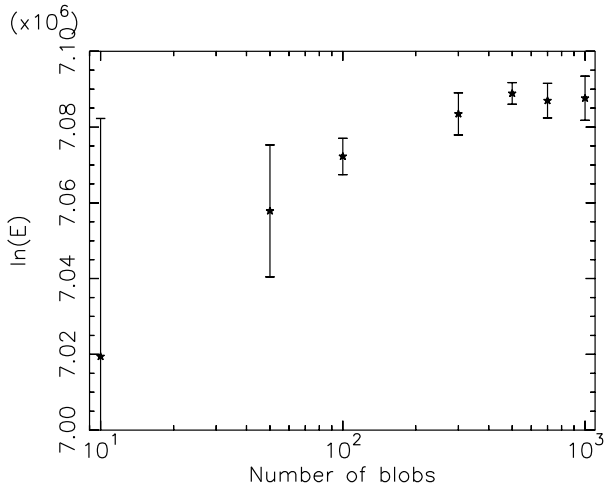} & 
      \quad \\
    \end{tabular}
  \end{center}
  \caption{The logarithm of the evidence (section \ref{sect:evidence})  calculated using
  the first 200 iterations for the simulated EPIC model data. The
  calculation was done 5 each times for  10, 50, 100, 300, 500, 700 and
  1000 particles.  The average and standard deviation of the evidence 
  for the 5 runs is shown as stars with error bars. \label{fig:evidence}}
\end{figure*}

In both runs the number of plasma particles was set to 700, a number
chosen by a simple evidence argument (section
\ref{sect:evidence}).  The evidence was calculated using the first 200
iterations running the method on the simulated EPIC data.  Evidence
calculations were made 5 times each  for 10, 50, 100, 500, 700 and 1000
particles (Figure~\ref{fig:evidence}). We find that the evidence reaches a
plateau or at least starts to level off at about 700~particles.

Each plasma particle was assigned the same allowed range of spectral and
spatial parameters, over which the prior pdf was uniform.  

The absorption was allowed to vary in the range; 0 to $1 \times
10^{21}~\mathrm{cm^{-2}}$. The metal abundances with respect to Solar
was varied locally between $0.0$ and $1.0$ Solar and  the redshift of the
plasma was fixed at $z=0.2$.  The particle temperature was assigned a
uniform prior over the range \TkeV{0.5} to \TkeV{9.5} in the EPIC run 
and over the range \TkeV{0.5} to \TkeV{4.5} in the RGS run. 
The particles are
modelled spatially using Gaussians with width whose logarithm was
allowed to vary over the range $0.0$ to $5.0$~arcseconds. Particle 
positions were constrained to a square area of $600'' \times 600''$ 
centered on the double-cluster centroid. 

The absorption was set to be a global parameter: the X-ray absorption
was assumed to be Galactic only and approximately the same over the extent of
the cluster. Plasma temperature is a local parameter, meaning that there
can be as many temperature phases at a given iteration as there are
particles. 

\begin{figure*}[!htb]
  \begin{center}
    \begin{tabular}{cc}
      \includegraphics[width=2.5in,angle=-90]{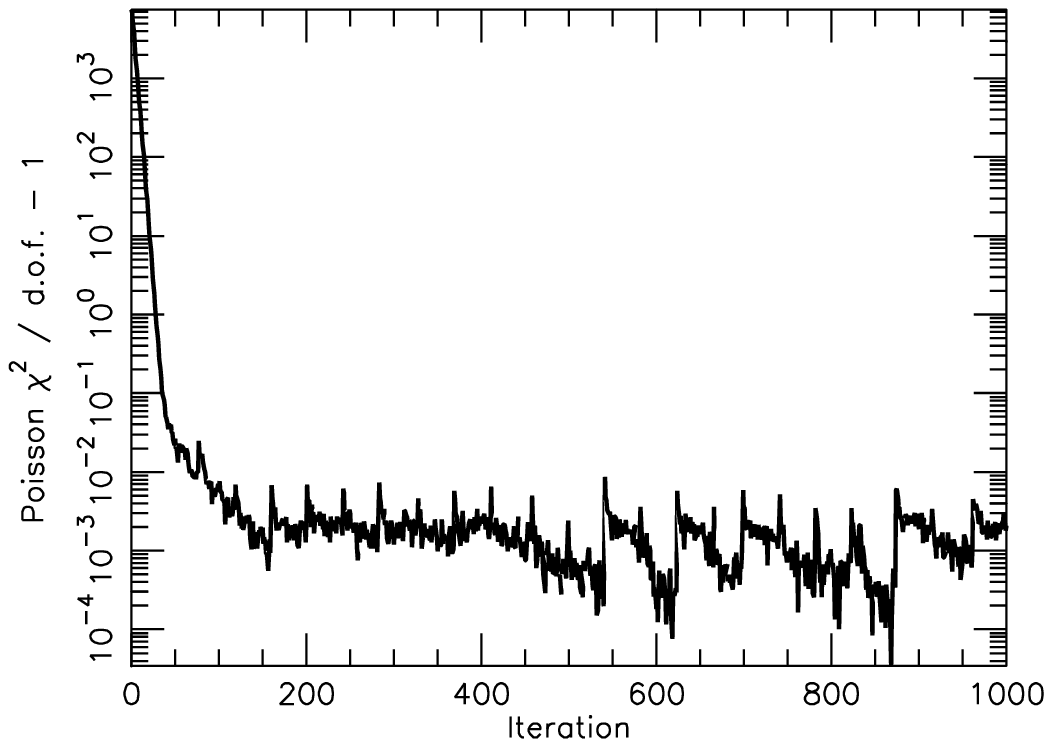} & 
      \includegraphics[width=2.5in,angle=-90]{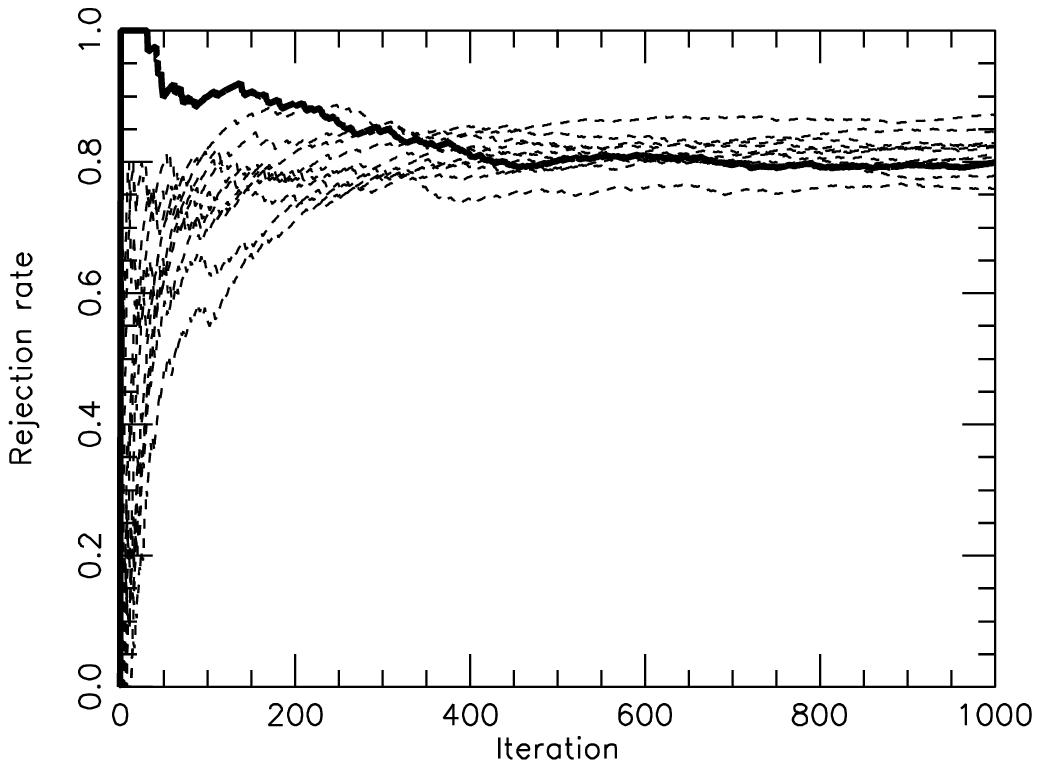} \\
    \end{tabular}
  \end{center}  
  \caption{The Poisson $\chi^2$ (see section \ref{sect:likelihood}) per degrees of freedom minus one (left) and rejection rate (right) for global (solid line) and a subset of local (dashed lines) parameters as a function of iteration number. \label{fig:statistics}}
\end{figure*}

Each iteration chain was run for 1000 iterations. This number of iterations is 
assumed to be sufficient to sample the posterior distribution to an acceptable 
level of accuracy. This choice is made to limit computing time to a 
practical level.
 
The evolution of the
Poisson $\chi^2$ per degrees of freedom minus one over the first  1000 iterations
is shown in Figure~\ref{fig:statistics} (left) along with the  rejection
rate (right). The rejection rate is shown for the global parameters 
(solid line) and local parameters for a sample of representative 10 particles (dashed 
lines). 

The renormalized log-likelihood can be seen to reduce to approximately 1
well before 200  iterations. Likewise, the rejection rate for both local
and global parameters reaches (and then slightly exceeds) the canonical
target value of 0.77 at approximately this iteration number. At this
point, and on average, any given  parameter has its value changed only
every fifth iteration.  

To ensure convergance of the chain we utilize the test suggested by 
Gelman in \citet{mcmc}.  This test involves running a number of 
chains with identical setups but different starting points.  We show as an
example 5 chains with the RGS version of the setup above for the column
density parameter.
Convergence is reached when a parameter $\phi_{i j}$ for chain 
$i = 1, ..., m$ and iteration $j = 1, ..., n$ has a variance 
between chains $B$ similar to the variance within chains $W$:

\begin{figure*}[!htb]
 \begin{center}
  \includegraphics[width=2.5in,angle=-90]{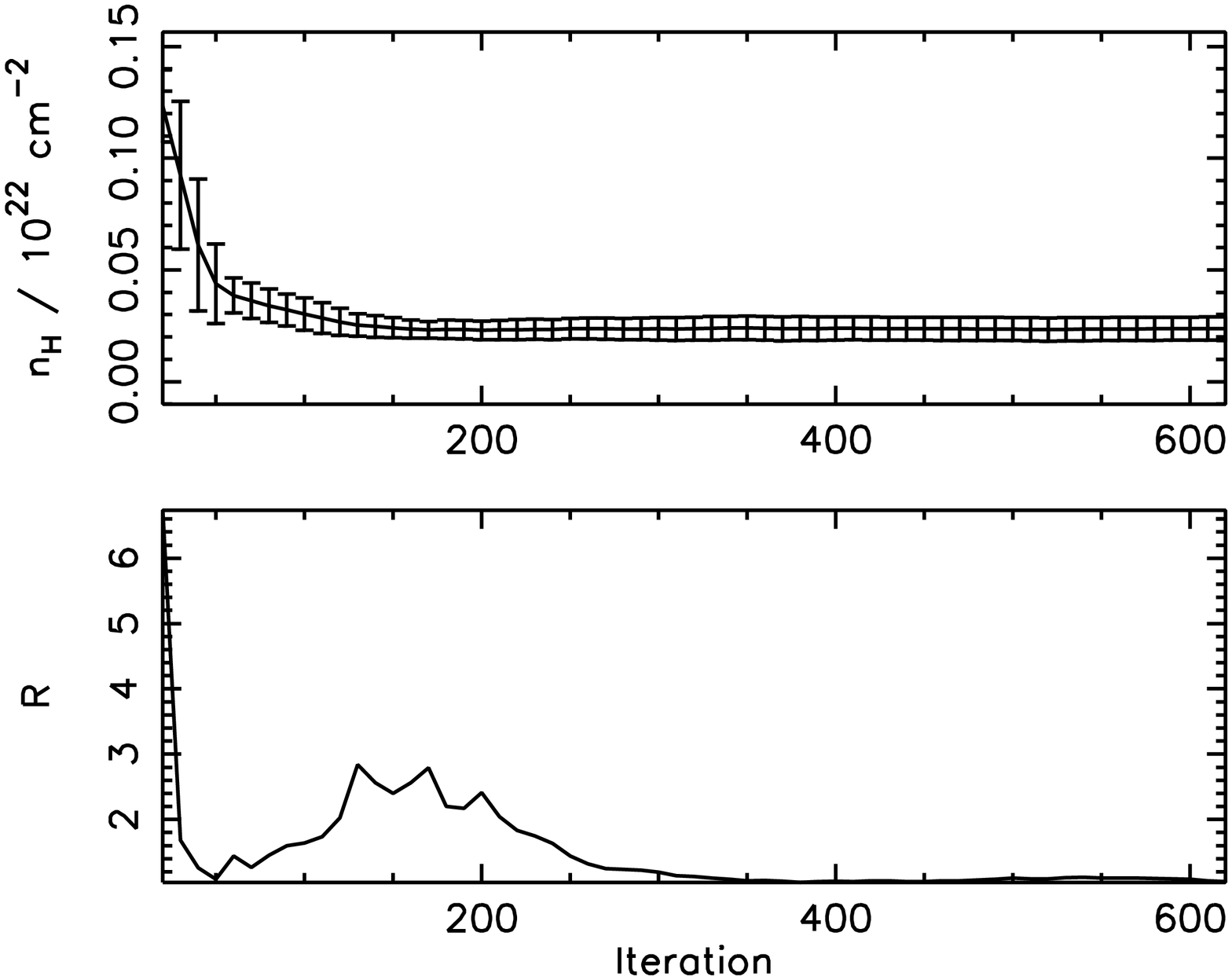}
 \end{center}
 \caption{Evolution of the $n_H$ parameter in the Gelman convergence test. 
The figure shows $\bar{\phi}$ with $\sqrt{\widehat{\mathrm{var}}\left(\phi\right)}$ 
errors (top).  The estimated potential scale reduction $\widehat{R}$ is shown below. 
\label{fig:convergence}}
\end{figure*}

\begin{equation}
B=\frac{n}{m-1} \sum^m_{i=1} \left( \bar{\phi_i} - \bar{\phi} \right) ^ 2
\end{equation}

\begin{equation}
W=\frac{1}{m} \sum^m_{i=1} s_i^2
\end{equation}

\noindent
where $\bar{\phi_i} = \frac{1}{n} \sum^n_{j=1} \phi_{i
  j}, \quad \bar{\phi} = \frac{1}{m} \sum^m_{i=1} \bar{\phi_i}, \quad s_i^2 =
\frac{1}{n-1} \sum^n_{j=1} \left( \phi_{i j} - \bar{\phi_i} \right)$.   By these variances, an overestimate of the variance, $\widehat{\mathrm{var}}\left(\phi\right)$ 
is formed 
\begin{equation}
\widehat{\mathrm{var}}\left(\phi\right) = \frac{n-1}{n} W + \frac{1}{n} B .
\end{equation}
The underestimate of the variance $W$ and $\widehat{\mathrm{var}}\left(\phi\right)$ 
will approach the true variance from opposite directions as $n \to \infty$.
Convergence is established when the 'estimated potential scale reduction' $\widehat{R}$:
\begin{equation}
\sqrt{\widehat{R}} = \sqrt{\frac{\widehat{var}\left(\phi\right)}{W}}
\end{equation}
is below 1.1.

In calculating the above variances and $\widehat{R}$ at iteration $i$ we 
discard the first $i/2$ iterations.  In our sample of chains the result for 
$n_H$ is shown in Figure~\ref{fig:convergence} with $\bar{\phi}$ shown with 
$\sqrt{\widehat{\mathrm{var}}\left(\phi\right)}$ errors (top) and $\widehat{R}$ 
(bottom). A value of $\widehat{R}$ below 1.1 was reached at iteration $\sim 400$. 

Since the Gelman test uses only iterations from $i/2$ to $i$ and convergence is 
reached at iteration $i=400$ the test shows that chains are more or less identical 
over the range $200$ to $400$. Therefore the chain was assumed to have reached
a stationary distribution after 200  iterations,  and consequently only
the last 800 iterations were used to infer the  properties of the
model. 


\subsection{Marginalized Distributions}
\label{sect:distributions}

The marginalized posterior distributions for a parameter can be
estimated simply by constructing the histogram of the sample parameter
values. One can use either all smoothed particles, or some suitable
subset allowing the cluster to be dissected in many different ways.  The
particles can for instance be selected for a certain range of sky
positions in order to view the distribution of temperatures in 
different parts of the cluster.  The width of the one-dimensional
histogram gives an estimate of the uncertainty on that one parameter,
with the caveat that the fluctuation in our Monte Carlo is likely to broaden
the distribution by approximately the factor, $\sigma_{L_{2N}}$, as in
the previous section.

\begin{figure*}[!htb]
  \begin{center}
    \begin{tabular}{cc}
      \includegraphics[width=2.5in,angle=-90]{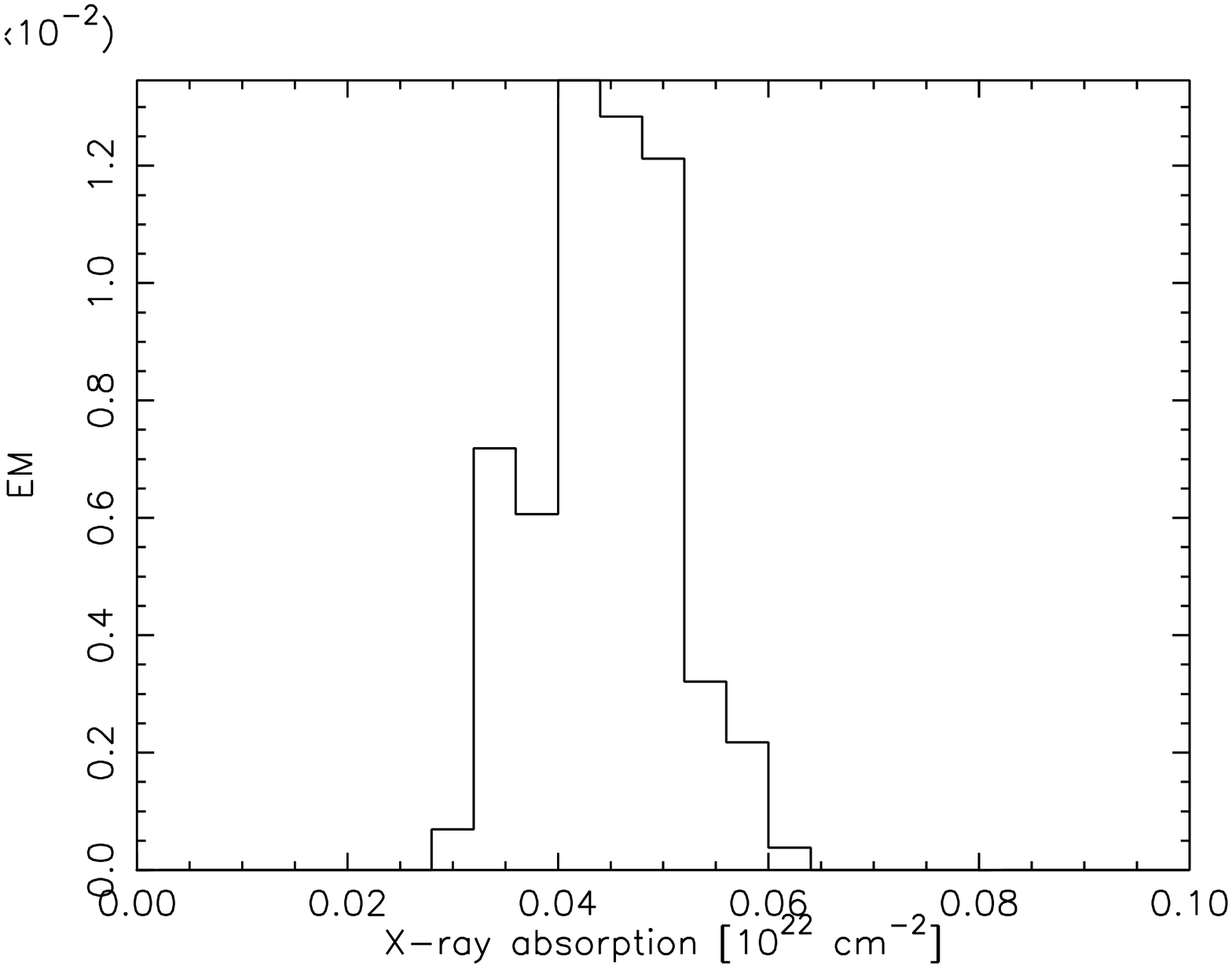} & 
      \includegraphics[width=2.5in,angle=-90]{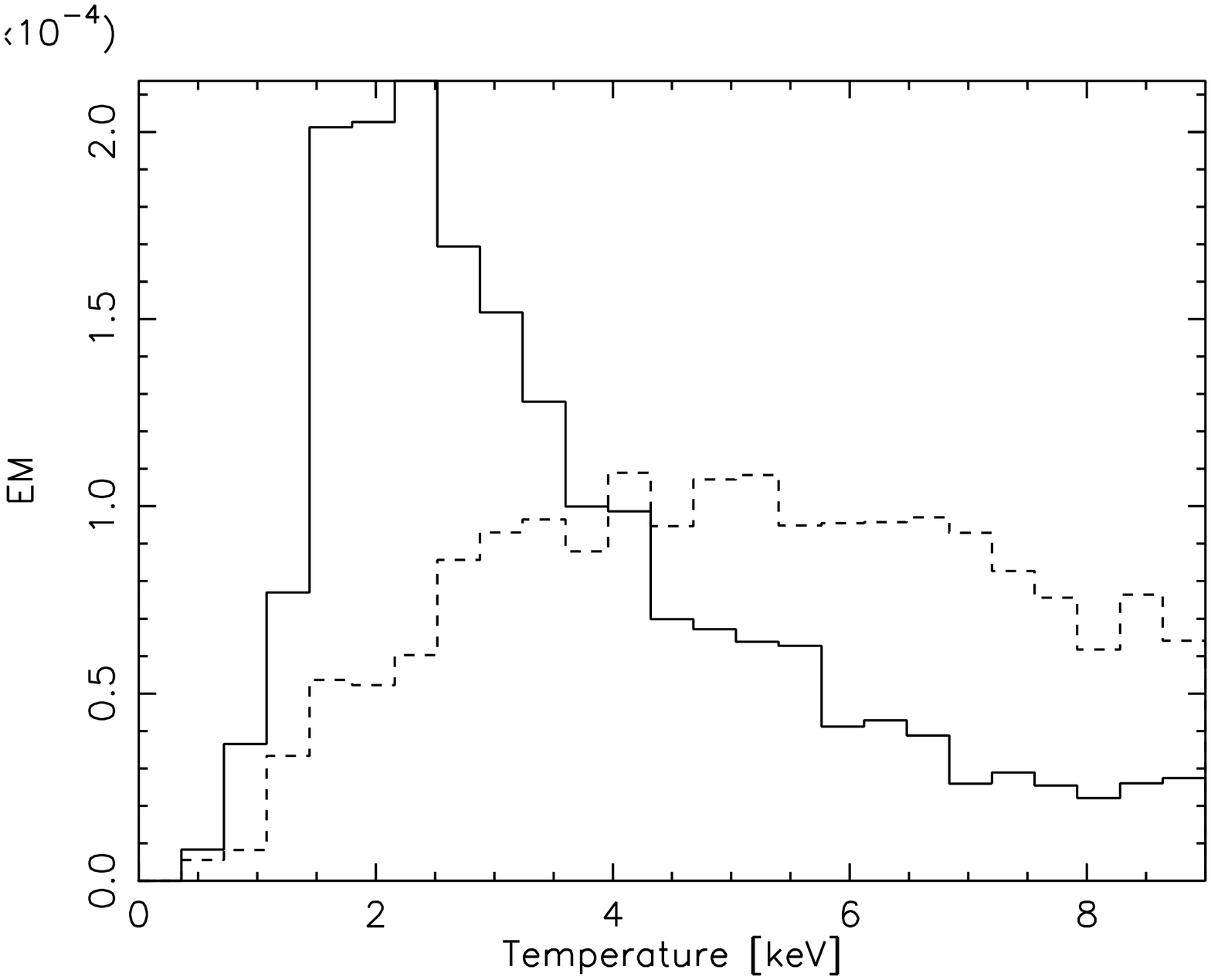} \\
      \includegraphics[width=2.5in,angle=-90]{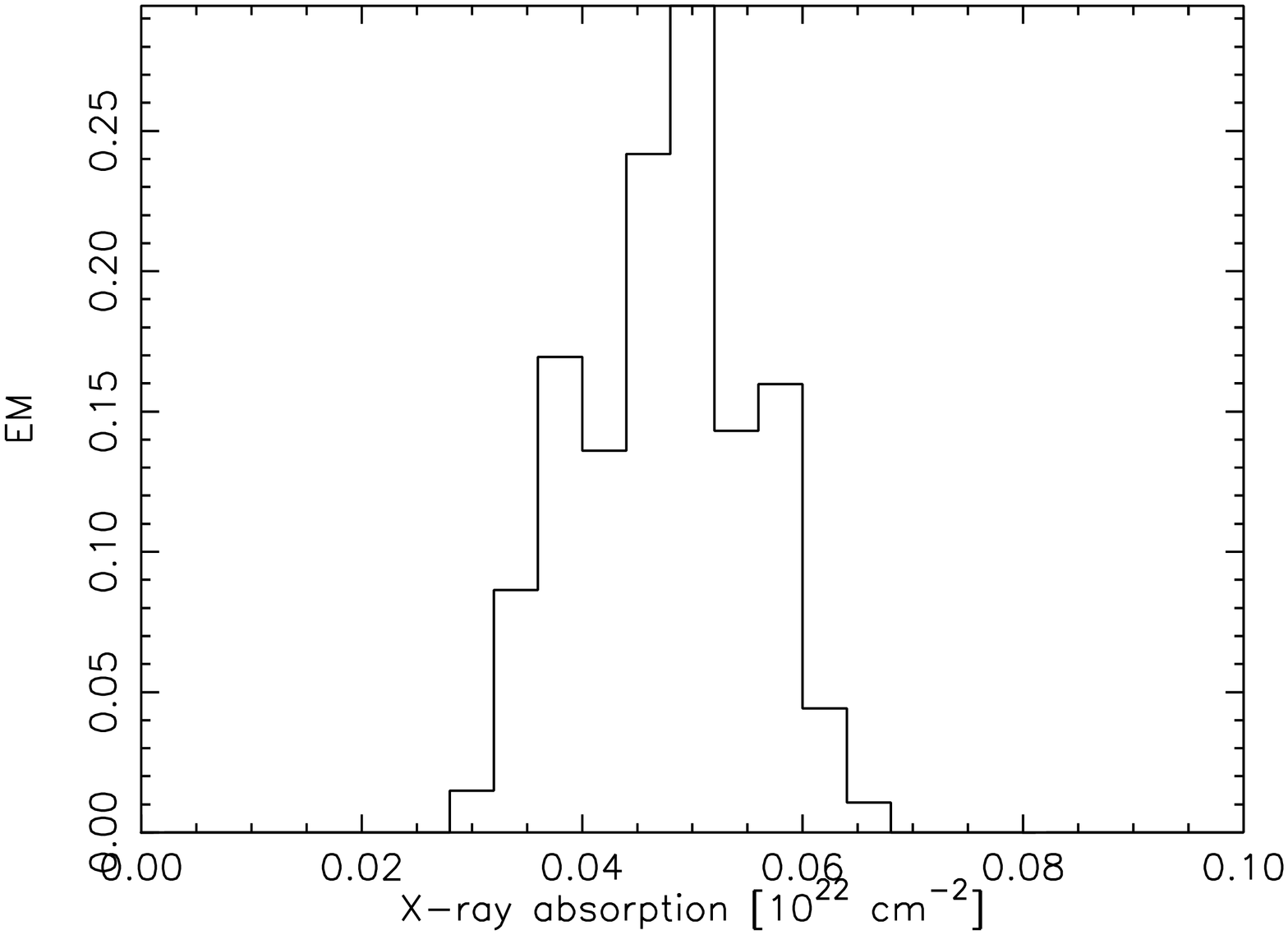} & 
      \includegraphics[width=2.5in,angle=-90]{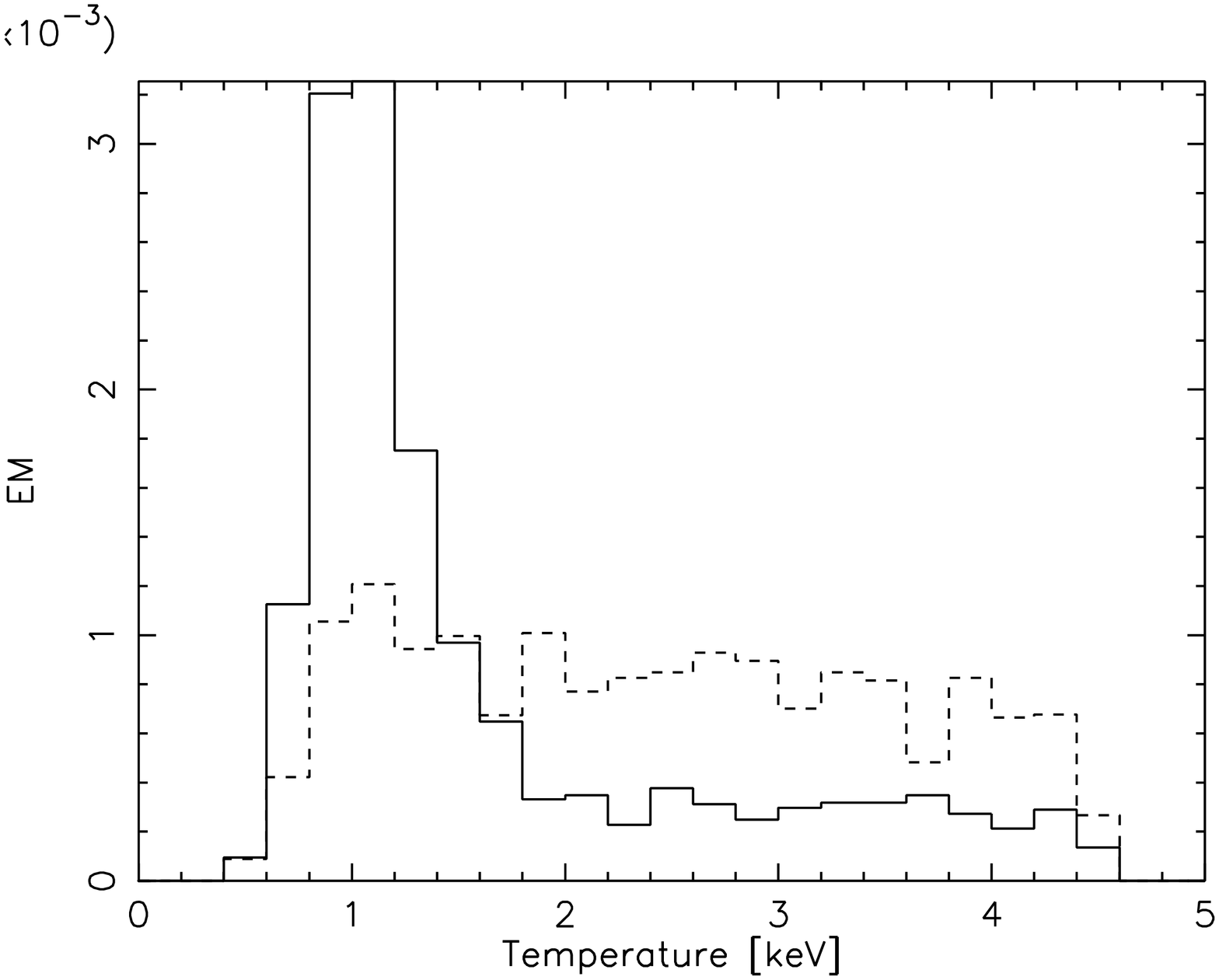} \\
    \end{tabular}
  \end{center}
  
  \caption{Parameter distributions for X-ray absorption (left) and ICM
temperature (right). The absorption is a global parameter and equal for
all particles while the temperature is a local parameter with different
values for all particles. The top panel shows the results of the EPIC run
and the lower panel contains the RGS run. For the plots of ICM 
temperature the blobs were selected in $20'' \times 20''$ boxes
centered  on the cold (solid line) and hot (dashed line) clusters.
\label{fig:histopars}}

\end{figure*}

For global parameters, like the X-ray absorption in this case, these 
distributions are straight-forward to interpret (see e.g.\
Figure~\ref{fig:histopars}, left panel). However, for local parameters
such as the plasma temperature these histograms will contain the
distribution of that parameter over all particles in addition to the local
uncertainty and Monte-Carlo noise.

 \begin{figure*}[!htb]
   \begin{center}
     \begin{tabular}{cc}
       \includegraphics[width=2.5in,angle=-90]{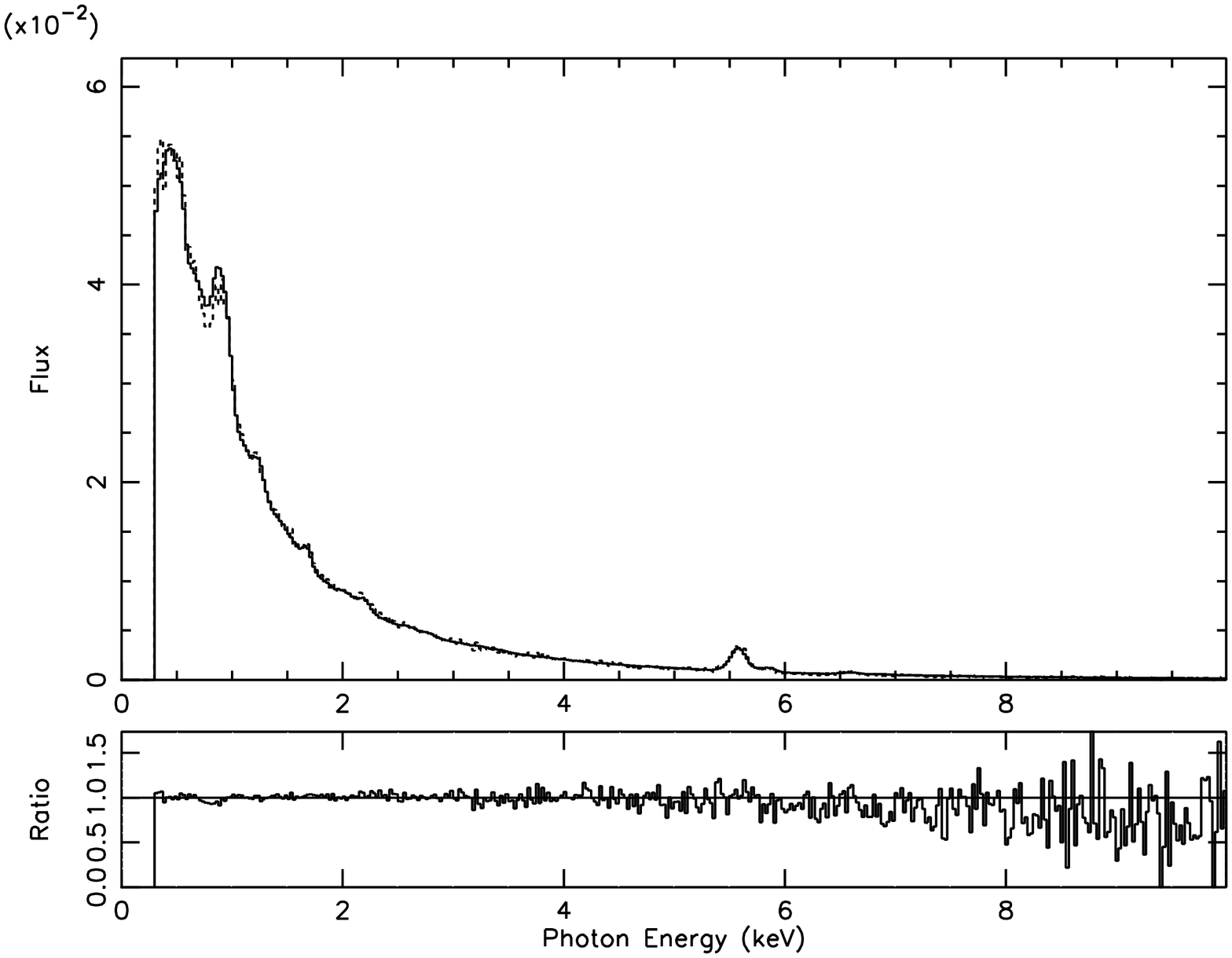} & 
       \includegraphics[width=2.5in,angle=-90]{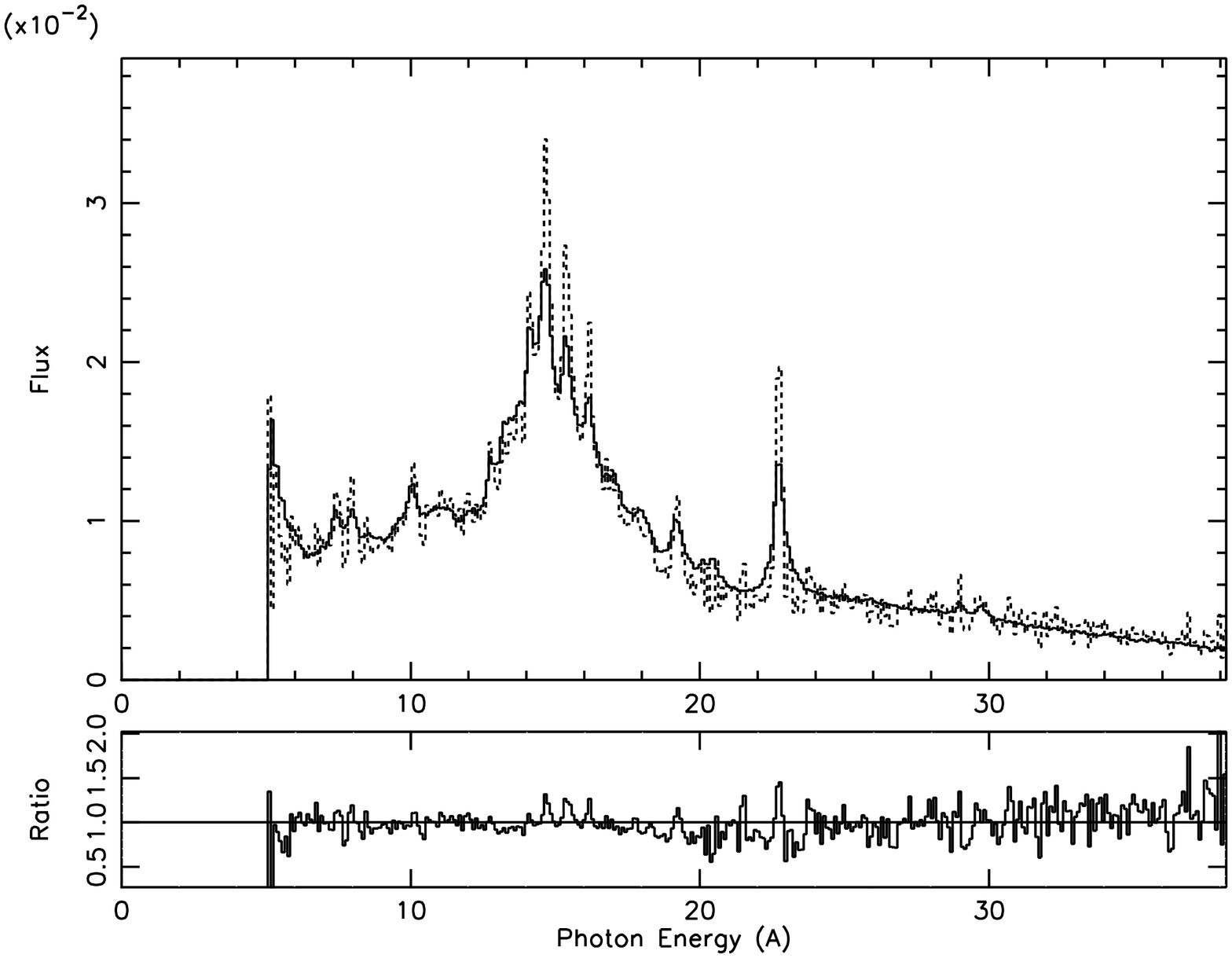} \\
     \end{tabular}
   \end{center}  
   \caption{Comparison of the fluxed spectra of the data (dashed) and model (solid) using models from iteration 200 to 1000. The flux (photons cm$^{-2}$ s$^{-1}$ keV$^{-1}$)) for the EPIC run is shown on the left and for the RGS run (photons cm$^{-2}$ s$^{-1}$ A$^{-1}$) on the right.  \label{fig:spectra}}
 \end{figure*}

In Figure \ref{fig:histopars} (left panel) we show the distribution of
the  X-ray absorption in units of $10^{22}~\mathrm{cm^{-2}}$ for the
last 800  iterations of the EPIC (top) and RGS (bottom) runs. 
Both pdf's comfortably accommodate
the true value of $0.04$.

The right panel of Figure \ref{fig:histopars} shows the distribution of
plasma temperatures for all particles over the last 800 iterations for the
EPIC (top) and RGS (bottom) runs. In order to view the separate 
distributions for the simulated clusters the plasma blobs were selected 
in $20'' \times 20''$ square regions centered on each cluster. The 
figure shows the distributions of the cold (lower left) and hot 
(upper right) components as solid and dashed lines respectively.
 The distribution
from the EPIC run shows the difficulty of determining plasma
temperatures from X-ray spectra. There is a high level of  degeneracy
between neighboring temperatures. We have allowed for 700 different
phases at any given iteration, making it possible (and indeed very
probable) that the  data from a single phase plasma will be
satisfactorily fit (and so returned by the sampler) with a wide range of
phases.  The peaks at \TkeV{3} and \TkeV{6} are not individually distinguishable
but are both broad distributions with significant probability mass
extending over the entire allowed interval. The prior has been updated
by the data though: the temperature distributions are not uniform, and
do actually contain useful information on the cluster temperature
structure. In the next section we derive a spatially-varying measure of
temperature from the median of this distribution at all spatial
positions, and show how this process recovers the input temperatures. 

In the RGS run the \TkeV{1} peak is reproduced more precisely whereas the \TkeV{3}
component  is represented as a combination of particles at approximately 
\TkeV{2-4.5}.  The RGS detectors lack sensitivity above energies of 2.25
keV, making them  makes them unresponsive to the hot cluster plasma. 
At low energies though, the high spectral resolution of RGS makes it an
excellent instrument for measuring cluster temperature distributions.


\subsection{Parameter Sky Maps} 

Spatial maps of quantities can be constructed by a number of methods.  
Working with particles in a luminosity basis, we can reconstruct the
surface brightness of the cluster by computing

\begin{equation}
\label{eqn:luminosity}
L(x,y) = \sum_i L_i \frac{e^{-\frac{\left({{\bf
          \rho}-{\bf \rho_i}}\right)^2}{2 \sigma_i^2}}}
          { 2 \pi \sigma_i^2 }
\end{equation}

\noindent where $L_i$ is the luminosity normalization for the $i$th
particle, $\sigma_i$ is the spatial width for the $i$th particle, and ${\bf
\rho}_i$ is the projected spatial position. Averaging the maps over the
samples then provides an image representing the particles' posterior PDF. Maps
generated in this way are, to within the precision given by the noise
and the accuracy dictated by the SPI model, corrected for  detector
exposure, vignetting, and PSF convolution, all of which  distort the raw
counts image. Furthermore, the particles are best-constrained where the
signal-to-noise is highest: the SPI average intensity maps are
multi-resolution image reconstructions.  

Likewise, luminosity-weighted parameter maps (e.g. temperature or
abundance maps) can be calculated by computing,  

\begin{equation}
\label{eqn:temperature}
T(x,y) = \frac{\sum_i L_i T_i \frac{e^{-\frac{\left({{\bf \rho}-{\bf \rho_i}}\right)^2}{2 \sigma_i^2}}}
          {2 \pi \sigma_i^2}}
          {L(x,y)}
\end{equation}

\noindent where $T_i$ is the value of the parameter (e.g temperature or
abundance) for  the $i$th particle, $L_i$ is the luminosity normalization,
$\sigma_i$ is the  Gaussian width, and ${\bf \rho}_i$ is the spatial
position.  

Note that the individual sample maps, while consistent with both the 
data and the noisy simulation, may not provide a reconstruction of the
cluster satisfactory to the eye of the analyst. For this, some sort of
averaging is required, preferably over a large set of samples (just as
when dealing with any uncertain measurements).  Such maps are discussed
in the following section.


\subsection{Median and Percentile Maps}

\begin{figure*}[!htb]
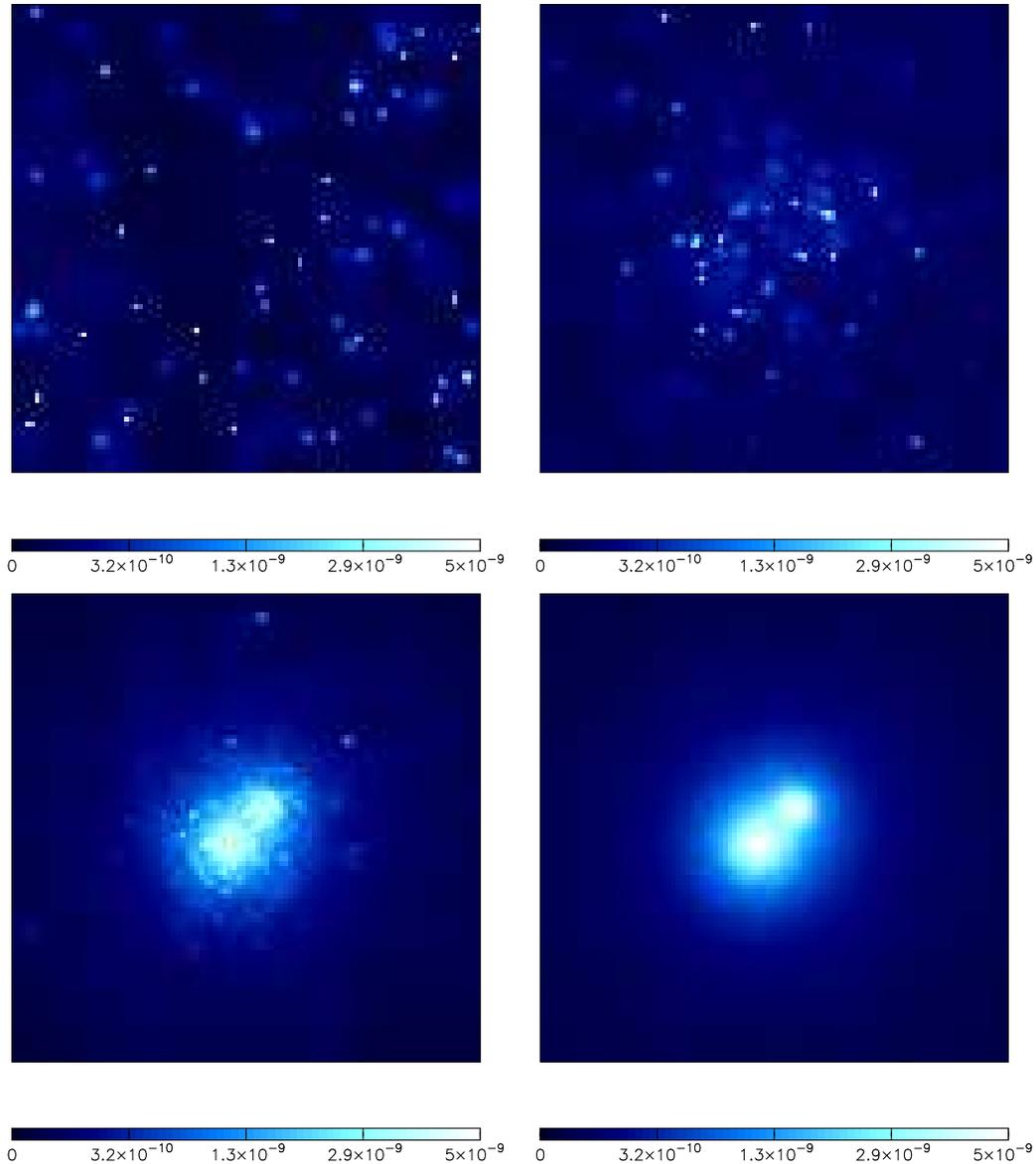

  \begin{center}
    \begin{tabular}{cc}
      \includegraphics[width=3in,angle=-90]{f7a.eps} & 
      \includegraphics[width=3in,angle=-90]{f7b.eps} \\
      \includegraphics[width=3in,angle=-90]{f7c.eps} & 
      \includegraphics[width=3in,angle=-90]{f7d.eps} \\
    \end{tabular}
  \end{center}
  
  \caption{$500'' \times 500''$ sky-maps of ICM bolometric luminosity (arbitrary units) 
    after 1, 10, 100 and 1000 iterations (top left to bottom right) 
    centered on the double-cluster 
    centroid. The luminosity is calculated according to Eq. 
    \ref{eqn:luminosity} and the median over all iterations is 
    taken in each $5'' \times 5''$ pixel. The maps are based on the 
    data from the EPIC run.\label{fig:lummed}}

\end{figure*}

\begin{figure*}[!htb]
  \begin{center}
    \begin{tabular}{cc}
      \includegraphics[width=2in,angle=-90]{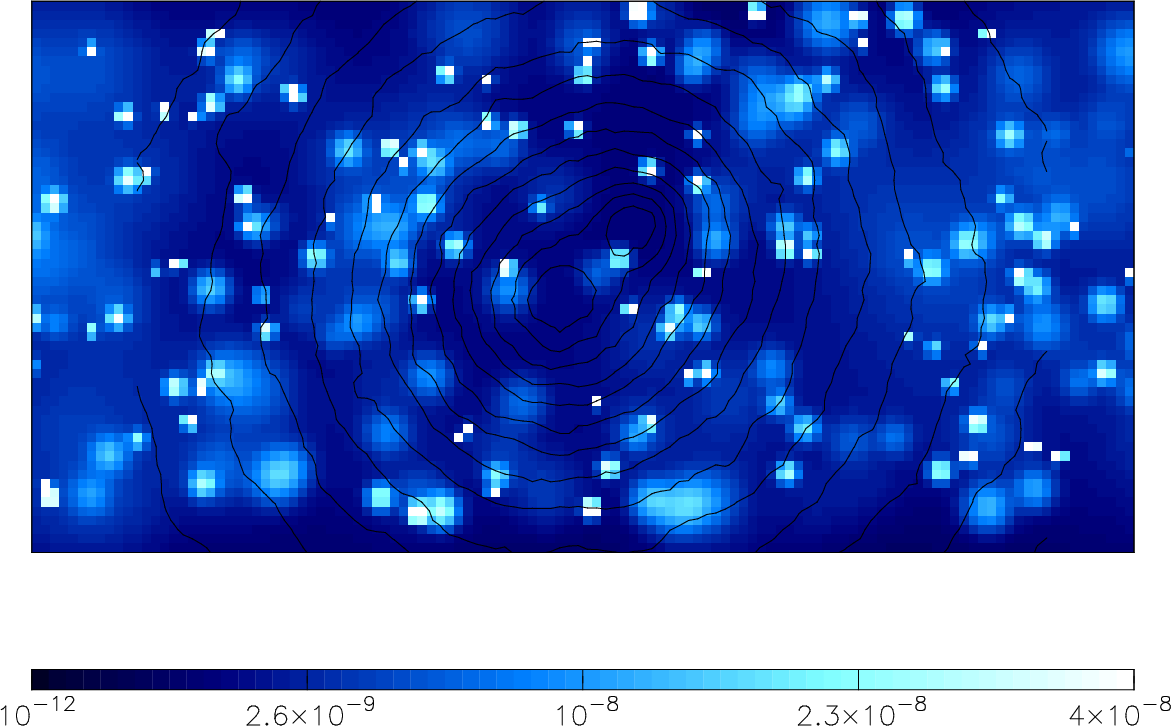} &
      \includegraphics[width=2in,angle=-90]{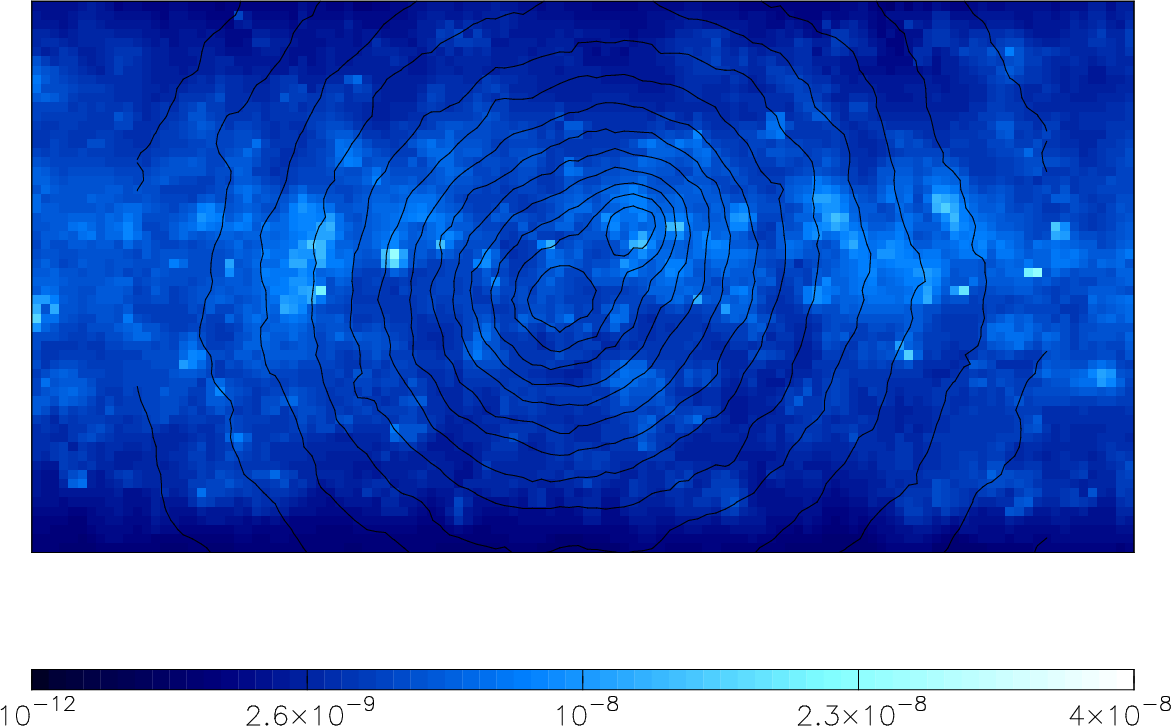} \\
      \includegraphics[width=2in,angle=-90]{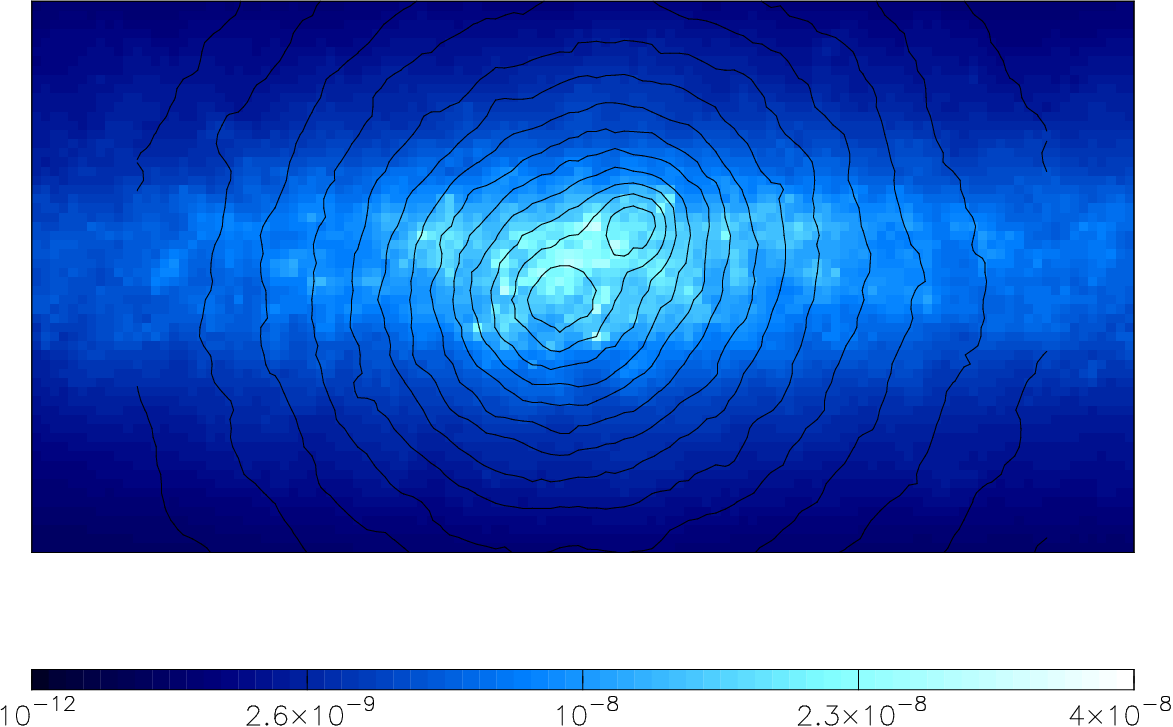} & 
      \includegraphics[width=2in,angle=-90]{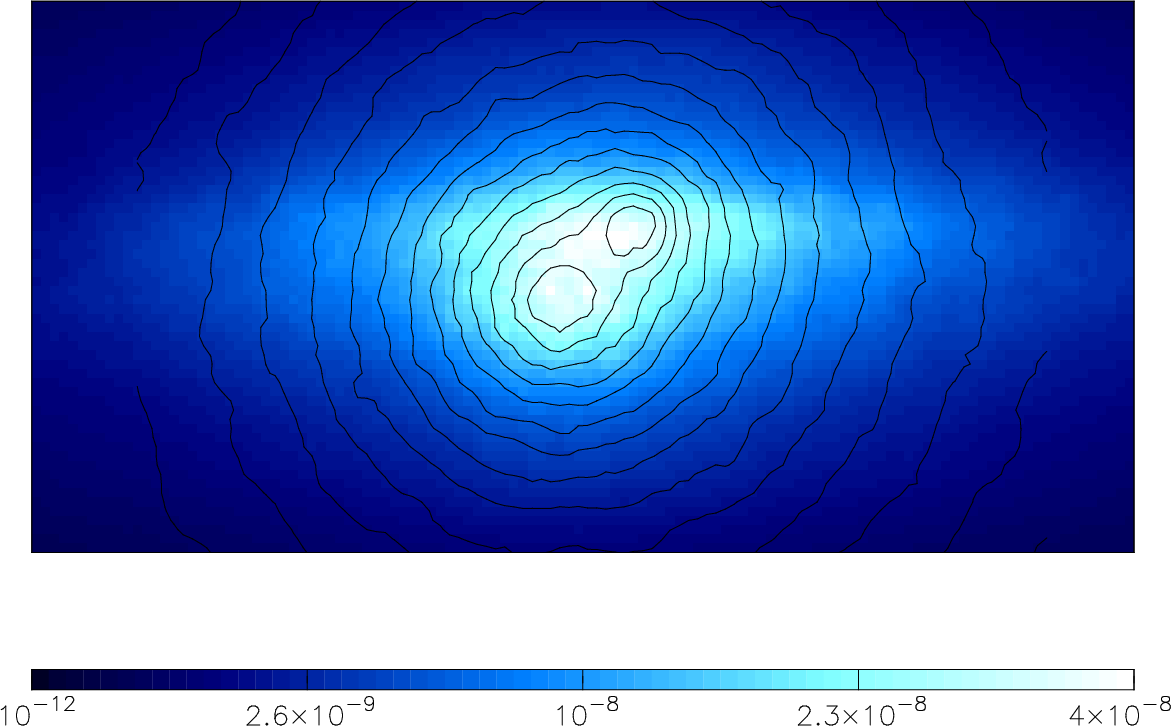} \\
    \end{tabular}
  \end{center}
  
  \caption{$600'' \times 300''$ sky-maps of ICM luminosity after 1, 10, 100, 
1000 iterations (top left to bottom right) for the RGS run. 
The contours show the luminosity obtained 
from the EPIC run after 1000 iterations for comparison. 
The particles are smeared due to
the uncertainty on the position in the dispersion direction of the
spectrometer. \label{fig:lummedRGS}}

\end{figure*}

The median map can be computed by taking the median 
of the $L_j(x,y)$ or $T_j(x,y)$ over all iterations $j$ in the 
stationary part of the Markov chain.
Median luminosity maps are shown in Figures \ref{fig:lummed} and 
\ref{fig:lummedRGS} for the EPIC and RGS runs respectively. 
As can be seen in the maps from the EPIC run the particles start out 
as a completely random distribution but after only 100 iterations 
the median map exhibits clear double cluster structure. The median 
luminosity map of the last 800 samples in the stationary part of 
the Markov chain is indistinguishible from the input model.
In the RGS run the clusters can be seen to be resolved 
approximately in the cross-dispersion direction after 1000 iterations. 

\begin{figure*}[!htb]
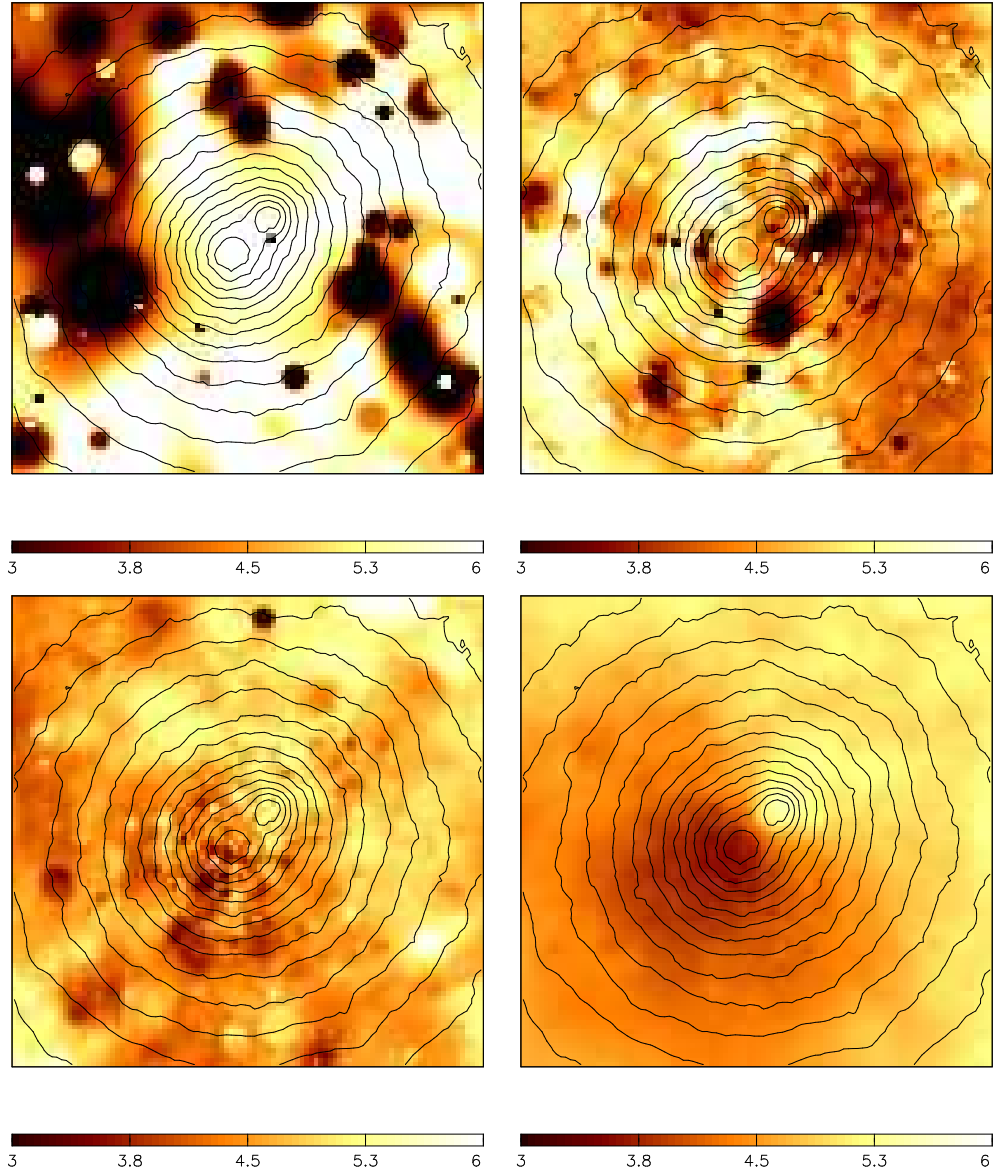

  \begin{center}
    \begin{tabular}{cc}
      \includegraphics[width=3in,angle=-90]{f9a.eps} & 
      \includegraphics[width=3in,angle=-90]{f9b.eps} \\
      \includegraphics[width=3in,angle=-90]{f9c.eps} & 
      \includegraphics[width=3in,angle=-90]{f9d.eps} \\
    \end{tabular}
  \end{center}
  
  \caption{$500'' \times 500''$ sky-maps of ICM temperature (keV) after 1, 
    10, 100 and 1000 iterations centered on the double-cluster 
    centroid. The temperature is calculated according to Eq. 
    \ref{eqn:temperature} and the median over all iterations is 
    taken in each $5'' \times 5''$ pixel. The contours show the 
    luminosity obtained from the EPIC run after 1000 iterations. 
    Based on data from the EPIC run. \label{fig:tempmed}}
\end{figure*}

Likewise, median maps of luminosity-weighted temperature are shown in
Figures \ref{fig:tempmed} and \ref{fig:tempmedRGS} for XMM-Newton EPIC
and RGS runs. It is clear from the sequence of plots in figure
\ref{fig:tempmed}  that it is more difficult to estimate spatially
resolved  cluster plasma temperatures than a luminosity distribution.
However, in the median map at iteration 1000 the two isothermal
components  are clearly  separated; even though the map is based on the
medians of very broad distributions (section \ref{sect:distributions}) 
we recover approximately the input temperatures of the two components.
Clearly, as seen in the maps based on the XMM-Newton RGS simulated
data (Figure \ref{fig:tempmedRGS}) creating spatially-resolved maps
using a spectrometer alone is more difficult; however, there is spatial
information there which is extracted with the SPI method. 

\begin{figure*}[!htb]
  \begin{center}
    \begin{tabular}{c}
      \includegraphics[width=3in,angle=-90]{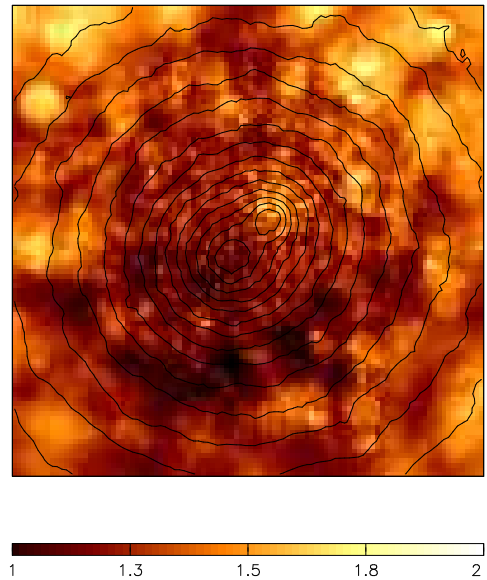} \\
    \end{tabular}
  \end{center}
  
  \caption{$500'' \times 500''$ sky-map, showing the width of the distribution for the median temperatures from iteration 200 to iteration 1000. The map depicts the difference of the 84th and 16th percentile of the distribution of the luminosity weighted temperatures for each iteration at any sky-coordinate.\label{fig:tempsig}}
\end{figure*}

Similarly, percentile maps can also be constructed instead of just the
median.  The difference between say the 84th percentile and 16th
percentile map gives the 1 $\sigma$ error map of a quantity in each
point ($x$,$y$)  with the caveat that the neighbouring pixels are highly
correlated due to the underlying smooth particle structure. This means
that the error map is an overestimate of the uncertainty on the pixel
values.  For a temperature map,  for example, this procedure gives the 1
$\sigma$ error of the luminosity weighted  temperature in each ($x$,$y$)
bin. Such a map, created from the 800 final samples in the  EPIC run, is
shown in Figure \ref{fig:tempsig}).  The map  displays a 1~$\sigma$
uncertainty of \TkeVsim{0.5} for the  bottom left (cold)
component and \TkeVsim{1} for the top  right (hot) component. 

This, again, demonstrates the subtle difficulties associated with
estimating high plasma temperatures given the currently available
instruments. That the error on the median appears to decrease to a small
value at spatial points far away from the cluster center is simply due
to the fact that for those distances the simulated data contain a very
small number of photons: any estimated temperature distribution will
default to the applied uniform prior. The median of this uniform
distribution will of course be close to the central value of the
range and not vary much in the evolution of the Markov chain. 

Of course, these percentile maps give only an estimate of the
uncertainty of the {\sl median} value. If we want to measure the width
of the intrinsic range of temperatures at any given sky coordinate,
should such a distribution exist, we have to do something more
sophisticated. One methods to visualize parameter ranges is 
demonstrated in the next section.

\begin{figure*}[!htb]
  \begin{center}
    \begin{tabular}{cc}
      \includegraphics[width=2in,angle=-90]{f10a.eps} & 
      \includegraphics[width=2in,angle=-90]{f10b.eps} \\
      \includegraphics[width=2in,angle=-90]{f10c.eps} & 
      \includegraphics[width=2in,angle=-90]{f10d.eps} \\
    \end{tabular}
  \end{center}
  
  \caption{$600'' \times 300''$ sky-maps of ICM temperature (keV) after 
1, 10, 100, 1000 iterations (top left to bottom right) for the RGS run. 
The contours show the luminosity obtained 
from the EPIC run after 1000 iterations for comparison. 
The particles are smeared due to
the uncertainty on the position in the dispersion direction of the
spectrometer. \label{fig:tempmedRGS}}
\end{figure*}

\subsection{Filtered Maps}

\begin{figure*}[!htb]
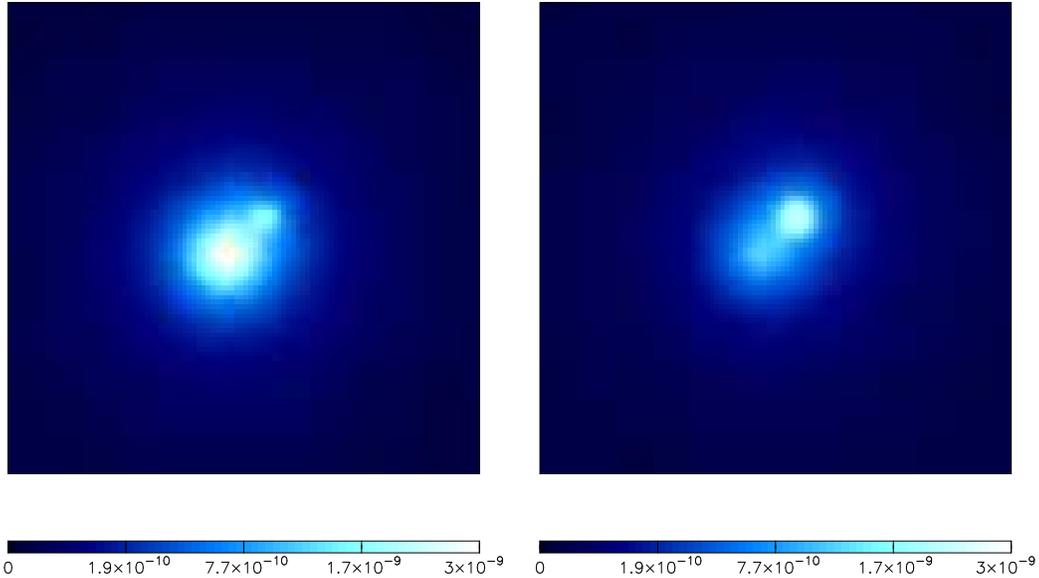

  \begin{center}
    \begin{tabular}{cc}
      \includegraphics[width=3in,angle=-90]{f12a.eps} &
      \includegraphics[width=3in,angle=-90]{f12b.eps} \\
    \end{tabular}
  \end{center}
  
  \caption{$500'' \times 500''$ sky-maps of filtered ICM luminosity 
    using particles of temperature \TkeV{0-4.5} (left) and 
    \TkeV{4.5-9} (right).\label{fig:lumfilt}}
\end{figure*}

Another useful way of examining the posterior pdf of a given
spatially-varying quantity is to segregate the particles into different
categories before constructing the luminosity maps.  A luminosity map
can be constructed from particles with temperatures between say
$1\times10^7~\mbox{K}$ and $2\times10^7~\mbox{K}$ (\TkeV{1-2}).   
This can then be
compared directly with other luminosity maps of other temperature ranges
to study the varying spatial distribution of the different plasma
temperatures.  To demonstrate this two filtered luminosity maps were
created, again using the final 800 samples of the EPIC chain
(Figure~\ref{fig:lumfilt}).  These maps were filtered for temperature
ranges \TkeV{0-4.5} (left)  and
\TkeV{4.5-9} (right) respectively. 
This type of luminosity map, filtered on plasma temperatures, would be 
impossible to produce using any standard spectral analysis: it requires 
some kind of forward-fitted spectral-spatial multicomponent model. 

\begin{figure*}[!htb]
  \begin{center}
    \begin{tabular}{cc}
      \includegraphics[width=2.5in,angle=0]{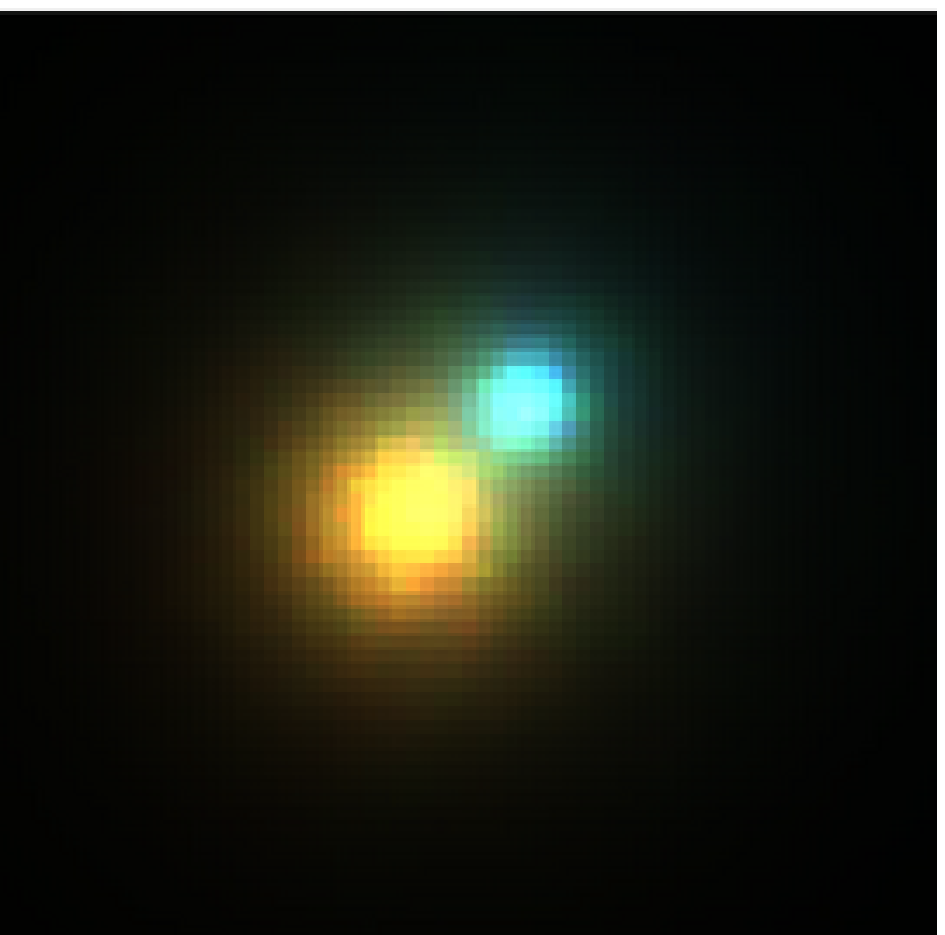} &
      \includegraphics[width=2.5in,angle=0]{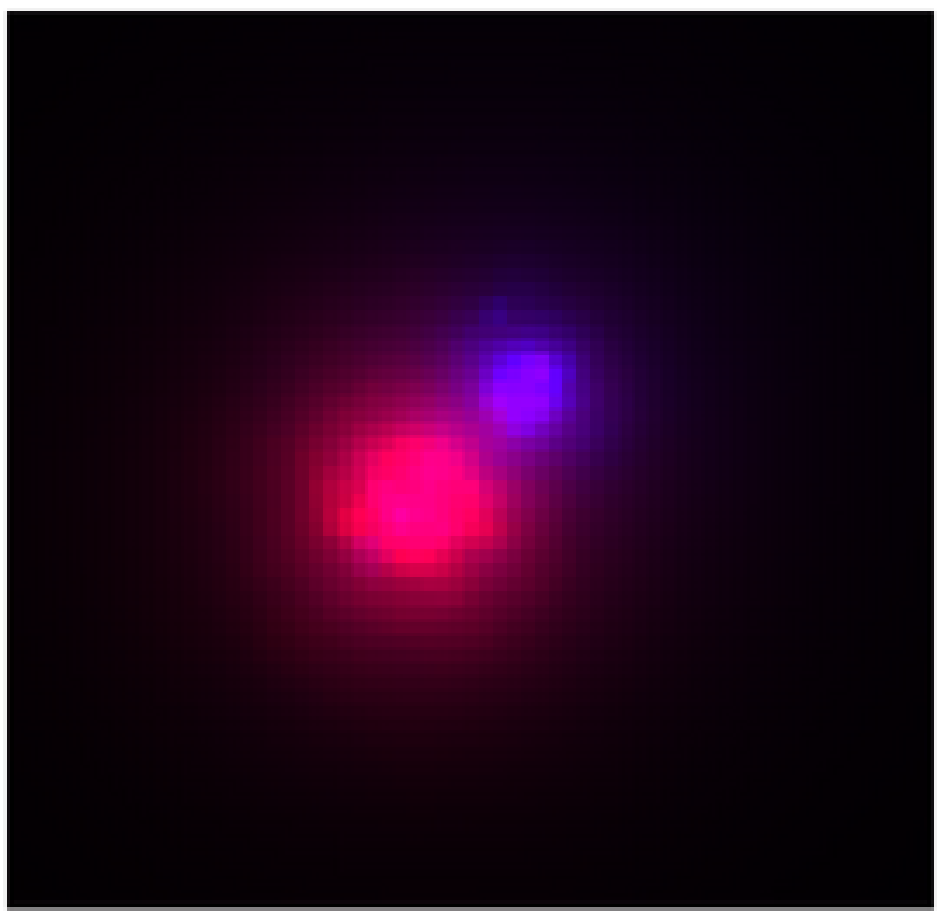} \\
    \end{tabular}
  \end{center}
  
  \caption{$300'' \times 300''$ RGB sky-map of ICM luminosity (left) where each color represents the luminosity of particles in different ICM temperature ranges. Here red represents particles with \TkeV{0-3}, green; \TkeV{3-6} and blue; \TkeV{6-9}. In the plot to the right the green component has been removed to make the difference of the hot and cold components clearer.\label{fig:rgb}}
\end{figure*}

Three-color RGB maps can be computed by using filtered luminosity maps 
with a different temperature range for each color. In
Figure~\ref{fig:rgb} such an RGB map is displayed using luminosity maps
filtered in three ranges  of temperature of equal size.  In principle,
this map contains more information since it describes the width of the
temperature distribution as well as the mean temperature.

\section{Conclusions and Future Work}

To summarize and conclude:

\begin{enumerate}
\item Motivated by the complexity of modern X-ray data on clusters of
galaxies, and the spatially- and spectrally-varying nature of both the 
emitting system and the instrument response, we have developed a highly
flexible forward-folding analysis scheme, dubbed smoothed particle
inference.
\item Deconvolution in the presence of high Poisson noise is achieved by
Monte Carlo simulation of mock data and comparison of this with
adaptively-binned raw counts with a two-sample likelihood function.
\item Clusters are modelled with a linear combination of Gaussian
emission profiles, with variable position, size, temperature and
abundance; these local parameters, and the global parameters associated
with the background emission and the galactic absorption, reside in a
highly multi-dimensional space which we explore with a Markov Chain
Monte Carlo sampler.
\item Testing this methodology on simulated data shows that the input
parameters can be recovered reliably, with precision limited by the shot
noise in the Monte Carlo mock data generation.
\item  SPI has been demonstrated to open up a variety of novel analysis
pathways: the translation of noisy data into uncertain particle parameters
brings the underlying physical distributions a significant distance
closer. 
\end{enumerate}

Multiple avenues can be taken in extending this work further.  Firstly,
since we have (iteratively) propagated a complex model through the XMM
instruments, it is straightforward to propagate the same model through
multiple instrument responses simultaneously.  Therefore, joint EPIC,
RGS,  Chandra, and Astro-E analyses, all of which would benefit from a 
careful treatment of the spatial-spectral response, may become routine.
Secondly, we anticipate that simultaneous X-ray, SZ and lensing analyses
could greatly benefit from a many-parametric approach such as this:
simpler analyses under-use the information in the X-ray data, allowing
biases to creep in.   Thirdly, we expect SPI analysis of low
signal-to-noise data sets, such as those obtained from X-ray cluster
surveys, to be very efficient: since the computation time scales with
the number of photons, analysis of simple data sets would be extremely
fast. The retention of maximum information in this data regime is
clearly of much interest.  

Fourthly, the particle positions can also be deprojected in similar ways to
standard deprojection techniques in order to model clusters in 3D: in a
future paper we will show how we do this using SPI, and  retaining the
full 2D cluster asymmetries.  Fifthly,  we anticipate that a variety of
hydrodynamical constraints could be applied to limit the vast parameter
space volume that is being considered.  To date we have used only
uninformative priors, as we attempt to move from data space to particle
space; the use of physical priors on the gas pressure should aid the
reconstruction of the cluster plasma distribution.  Finally, this work
can be seen as the beginnings of a  bridge to theoretical physical
models of cluster formation and evolution; one can envisage such models
being developed under constraints imposed by the data in something like
the methodolgy outlined here. Smoothed particles sit very well in the
hierarchical picture of cluster formation: we have good reason to
believe that they may be the optimal basis set with which to reconstruct
their images.  


\appendix
\section{Derivation of Likelihood Function}

The basic datum is a number of photons in a three dimensional bin $(u_1, u_2,
p)$. The bin size may be set by physical limits such as the extent of the CCD 
pixel, or be imposed given some understanding of, for example, the spectral
resolution of the instrument. The simulated photons are then binned on the same
grid. Assume that we have $n$ photons in the observed dataset, and have
simulated $m$ photons from our model: in the $i^{\rm th}$ bin we have  $E_i$
simulated photons, and $O_i$ observed photons.  Since the photons falling into
a set of bins will follow a multinomial distribution, we can use the usual
multinomial likelihood function

\begin{equation}
L_1 = n ! \prod_i \frac{{\left( \frac{E_i}{m} \right)} ^ {O_i}}{O_i !}
\end{equation}

\noindent 
where we have used $\frac{E_i}{m}$ as a proxy for the actual probability of
finding a photon in the $i$th bin.  In the limit of large m, this will
approach the converge to the true probability of the model.

A more sophisticated likelihood function is one that attempts to include the
error induced by the Monte Carlo noise on the true probability.  We then
calculate, $L_2$, which is the product of two likelihood functions for both
the detected and simulated photons (i.e. a two sample likelihood function)

\begin{equation}
L_2 = n ! m !  \prod_i \frac{{{\left( \frac{O_i+E_i}{m+n} \right)}}^{O_i} 
{{\left( \frac{O_i+E_i}{m+n} \right)}}^{E_i} }{O_i ! E_i!}
\end{equation}

\noindent
where we have used $\frac{O_i+E_i}{m+n}$ as an estimate for the probability in
the $i$th bin.  

Another approach would be to not assume that the probability, $p_i$ is
estimated by $\frac{O_i+E_i}{m+n}$, but instead integrate this probability
distribution over a prior distribution for that probability.  This is
expressed as, 

\begin{equation}
L_3 = n! m! \prod_i \int \frac{p_i^{O_i} p_i^{E_i}}{O_i ! E_i!} f(p_i) dp_i
\end{equation}

\noindent
where $f(p_i)$ is the prior distribution.  We have not, however, been able to obtain both an
analytic and properly normalized expression for $L_3$, in the case where
the numbers of simulated and observed photons are constrained to be the
same.  Therefore, we use
$L_2$ and note that in the limit of large $m$ these expressions will be
identical. This is indeed the case for unconstrained photon numbers
where the Poisson distribution is
appropriate for the 
likelihood of the counts in each bin.
Future study, however, may develop these statistics further.

It is also possible to normalize these likelihood functions, such that they
will approximately follow the $\chi^2$ distribution.  This is done by dividing
by the same expression with the probability estimates replaced by the
actual number of counts.  So the normalized expression for $L_2$ can be
written as

\begin{equation}
 L_1^N = \prod_i \frac{\frac{{\left( \frac{E_i}{m} \right)} ^ {O_i}}{O_i
     !}}{\frac{{\left(\frac{O_i}{n}\right)}^{O_i}}{O_i !}}
= \prod_i \frac{{\left( \frac{E_i}{m} \right)} ^ {O_i}}{{\left(\frac{O_i}{n}\right)}^{O_i}}
\end{equation}

\noindent
and for $L_2$ it is

\begin{equation}
L_{2N} = \prod_i \frac{\frac{{{\left( \frac{O_i+E_i}{m+n} \right)}}^{O_i} 
{{\left( \frac{O_i+E_i}{m+n} \right)}}^{E_i} }{O_i ! E_i!}}{\frac{{{\left( \frac{O_i}{n} \right)}}^{O_i} 
{{\left( \frac{E_i}{m} \right)}}^{E_i} }{O_i ! E_i!}}
= \prod_i \frac{{{\left( \frac{O_i+E_i}{m+n} \right)}}^{O_i+E_i} }
{{{\left( \frac{O_i}{n} \right)}}^{O_i} 
{{\left( \frac{E_i}{m} \right)}}^{E_i} }
\end{equation}

\noindent
Note that $-2 \log{L_1^N}$ and $-2 \log{L_{2N}}$ will be distributed as $\chi^2$ with the
number of degrees of freedom equal to the number of bins.  We use $L_{2N}$ for
all future analysis.  In the appendix we demonstrate how this statistic
approaches the two sample $\chi^2$ statistic that we used in the large $m$ and
$n$ limit.

Following, we demonstrate how the likelihood function, we described in \S2.3
approach the two sample $\chi^2$ distribution.

\begin{equation}
L_{2N} = \prod_i \frac{{{\left( \frac{O_i+E_i}{m+n} \right)}}^{O_i+E_i} }
{{{\left( \frac{O_i}{n} \right)}}^{O_i} 
{{\left( \frac{E_i}{m} \right)}}^{E_i} }
\end{equation}

Note, if we take the logarithm of the above equation, we obtain

\begin{equation}
\log{L_{2N}} = \sum_i O_i \log{n\frac{\frac{O_i+E_i}{m+n}}{O_i}}
+E_i \log{m\frac{\frac{O_i+E_i}{m+n}}{E_i}}.
\end{equation}

Rewriting this,

\begin{equation} 
\log{L_{2N}} =\sum_i O_i
\log{\left(1+\frac{n\frac{O_i+E_i}{m+n}-O_i}{O_i}\right)}
\end{equation}

\begin{equation}
+E_i \log{\left(1+\frac{m\frac{O_i+E_i}{m+n}-E_i}{E_i}\right)}.
\end{equation}

Expanding these around $\frac{n\frac{O_i+E_i}{m+n}-O_i}{O_i} \equiv \epsilon_1
\approx 0$ and
$\frac{m\frac{O_i+E_i}{m+n}-E_i}{E_i} \equiv \epsilon_2 \approx 0$, we obtain

\begin{equation}
\log{L_{2N}} = \sum_i O_i \left( \epsilon_1 - \frac{\epsilon_1^2}{2} \right) +
E_i \left( \epsilon_2 - \frac{\epsilon_2^2}{2} \right)
\end{equation}

Exponentiating this equation gives,

\begin{equation}
L_{2N} = \prod_i e^{-\frac{\left( O_i- n \frac{O_i+E_i}{m+n} \right)^2}{2 O_i}}
e^{-\frac{\left( O_i- n \frac{O_i+E_i}{m+n} \right)^2}{2 O_i}} e^{n\frac{O_i+E_i}{m+n}-O_i}
e^{m\frac{O_i+E_i}{m+n}-E_i}
\end{equation}

The last two factors multiplied together are one, so we obtain,

\begin{equation}
L_{2N} = \prod_i e^{-\frac{\left( O_i- n \frac{O_i+E_i}{m+n} \right)^2}{2 O_i}}
e^{-\frac{\left( O_i- n \frac{O_i+E_i}{m+n} \right)^2}{2 E_i}}
\end{equation}

when both $m$ and $n$ are large.  For small $\epsilon_1$ and $\epsilon_2$, this
is also equivalent to

\begin{equation}
L_{2N} = \prod_i e^{-\frac{\left( O_i- n \frac{O_i+E_i}{m+n} \right)^2}{2 n \frac{O_i+E_i}{m+n}}}
e^{-\frac{\left( E_i- m \frac{O_i+E_i}{m+n} \right)^2}{2 m \frac{O_i+E_i}{m+n}}}
\end{equation}

and this is exactly equal to

\begin{equation}
L_{2N} = e^{-\frac{\chi^2_{2S} }{2}}.
\end{equation}

\noindent
where $\chi^2_{2S}$ is the two sample $\chi^2$ function that was used in 
\citet{peterson3}.  Its reduced value would be close to one for a good fit.


\appendix
\section{Test of Parameter Update Method}

In order to determine if the interation sequence we used in \S2.4.3 is valid
and approximately obeys detailed balance we performed several simulations.  We
use the double isothermal beta model in \S3.1 as the input simulation.  We
then fit this same model to the simulation instead of using the smoothed
particles.  In this way, the model is much simpler and only has the following
15 parameters, column density, the two temperatures, core radii,
$\beta$, the positions (4 parameters), the two abudances, and overall
normalizations.  With this few numbers of parameters, we can update the
parameters in three different ways.  First, we update the parameters treating
all the parameters as global parameters defined in \S2.4.3.  This is a proper
Markov chain obeying detailed balance and all parameters are modified
simulataneously in a single Markov step.  Second, we update the parameters for
one beta model as global parameters and the other parameters as local
parameters.  Third, we update all the parameters as local parameters.  The
second two approaches are not standard Markov chains, but essentially create
small Markov chains at each iteration for some of the parameters.  

\begin{figure*}[!htb]
  \begin{center}
    \begin{tabular}{cc}
      \includegraphics[width=2in,angle=-90]{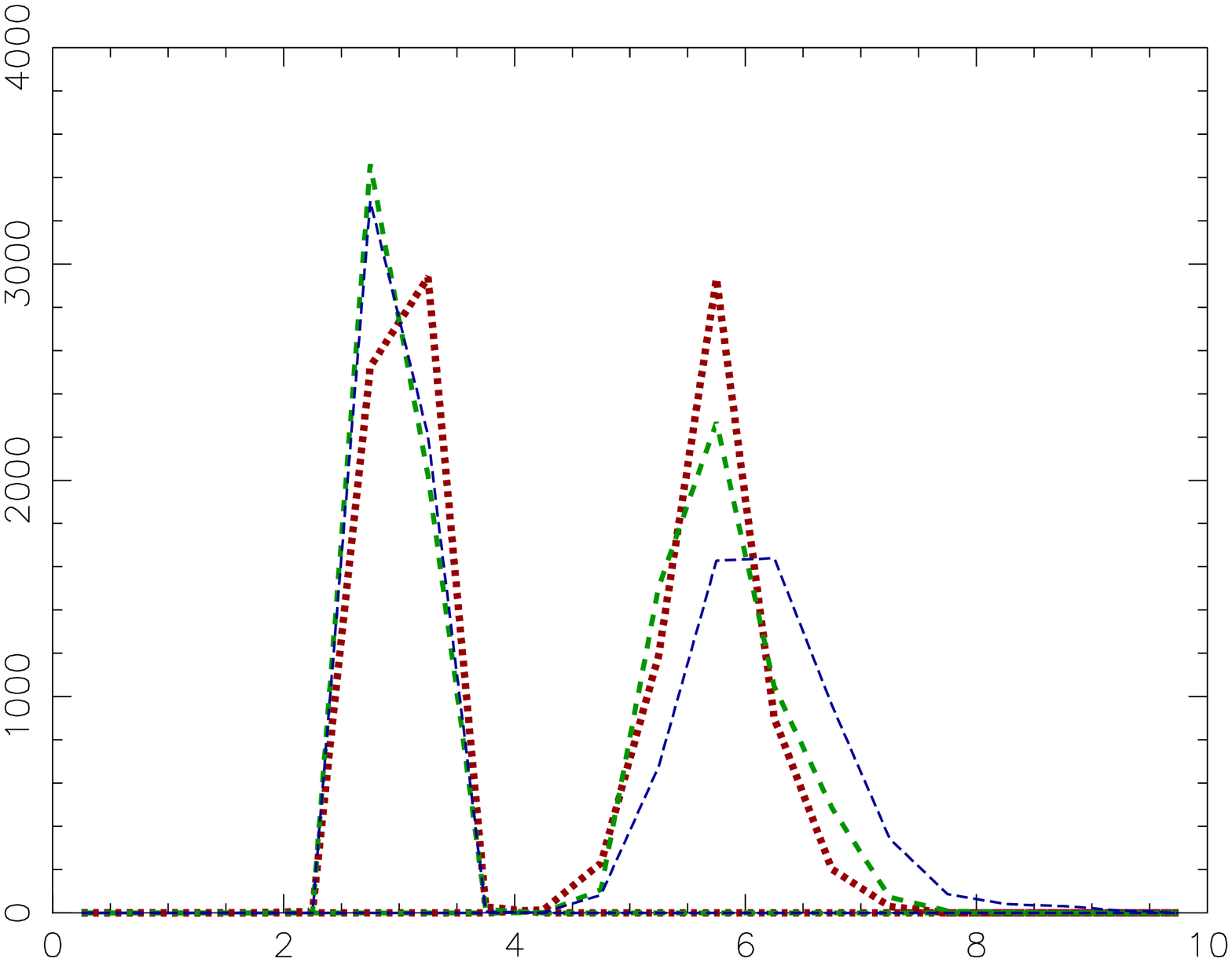} & 
      \includegraphics[width=2in,angle=-90]{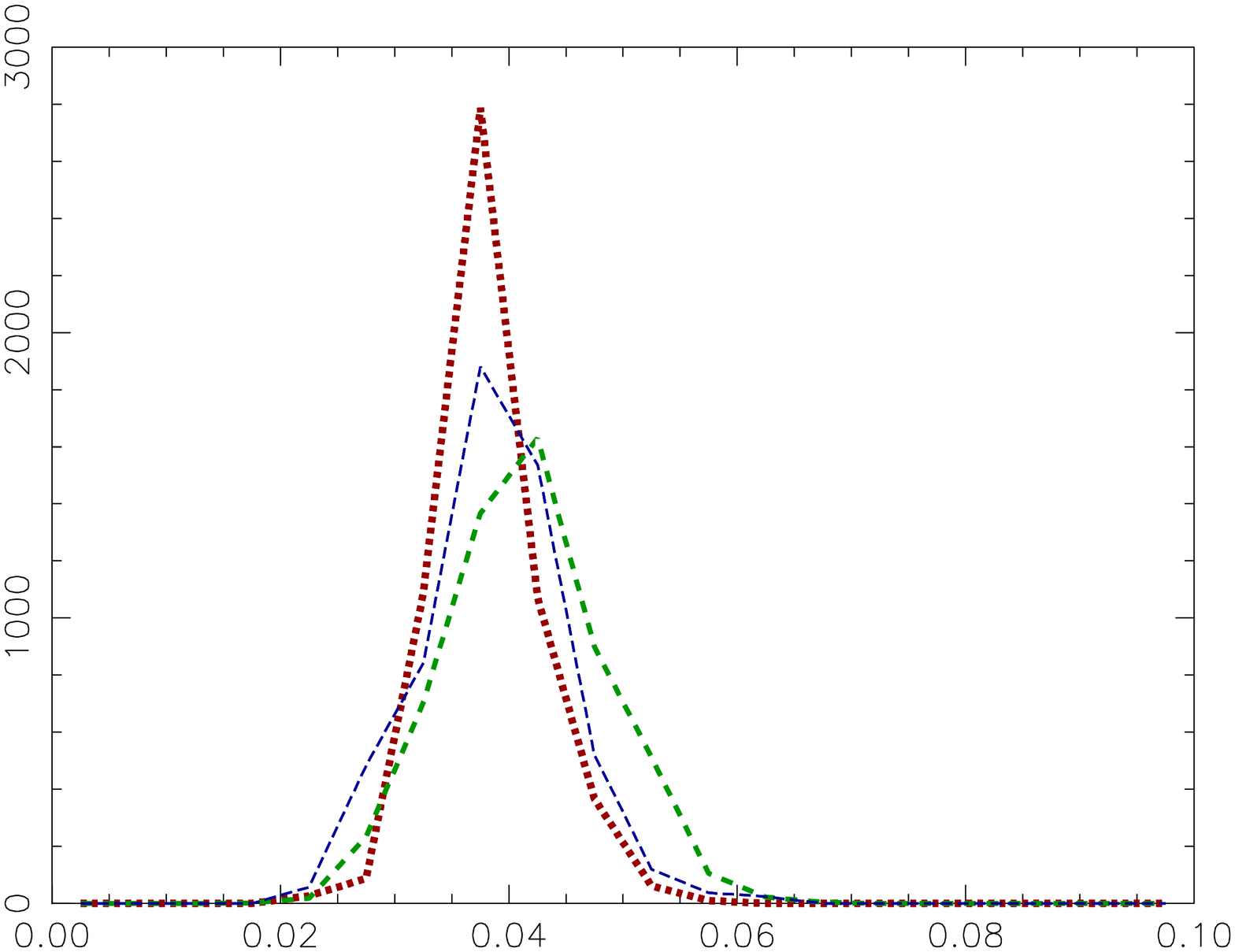} \\
    \end{tabular}
  \end{center}
  \caption{Posterior distributions of temperature (right) and absorption (left) using three different parameter update schemes; all global (dotted), mixed global and local (thin dashed) and all local parameters (thick dashed). \label{fig:globallocal}}
\end{figure*}

We display the posterior distributions of the temperature and absorption
parameters in Figure~\ref{fig:globallocal}.  In all three simulations the posterior
distribution is identical.  This demonstrates that the various iteration scheme
approximately obey detailed balance.  This suggests that it is not critically
important what scheme we use and works in the limit that the posterior
distribution becomes stationary.  These simulations also demonstrate that the
posterior temperature distribution is smaller when the model is constrained to
only consist of two temperatures than the other simulations with smoothed
particles because of the simplicity of the model.  The posterior distribution
is smallest at low temperatures because of the nature of X-ray spectra.


\acknowledgments

We acknowledge many helpful discussions on the suitability of smooth
particles for modelling cluster data with Steve Allen, Ted Baltz, Roger
Blandford, Sarah Church, Steve Kahn, Greg Madejski, Frits Paerels, Vahe
Petrosian, Yoel Raphaeli, and Keith Thompson.  Likewise we thank Garrett
Jernigan, Mike Hobson, Sarah Bridle and John Skilling for similar input
on the statistical details of the method. Financial support for KA is
provided  by the G\"{o}ran Gustavsson Foundation for Research in Natural
Sciences  and Medicine.  This work was supported in part by a
grant from NASA for scientific and calibration support of the RGS at
Stanford.  This work was also supported in part by the U.S.
Department of Energy under contract number DE-AC02-76SF00515.




\begin{thebibliography}{}



\bibitem[{Andersson \& Madejski}(2004)]{andersson} Andersson, K. \& Madejski, G. 2004, ApJ 607, 190.

\bibitem[{Andersson et al.\ }(2006)]{andersson2} Andersson, K. et al. 2006, in preparation.

\bibitem[{Arabadjis, Bautz \& Garmire}(2002)]{arabadjis} Arabadjis, J. S., Bautz, M. W. \& Garmire, G. P.  2002, ApJ, 572, 66

\bibitem[{Arnaud}(2001)]{arnaud} Arnaud, M. et al. 2001, A\&A 365, 80.

\bibitem[{Buote et al.}(2003)]{buote} Buote, D. A., Lewis, A. D., Brighenti, F., \& Mathews, W. G. 2003, ApJ 594, 741.

\bibitem[{Bonamente et al.}(2004)]{bonamente}  Bonamente, M., Joy, M.~K., Carlstrom, J.~E., Reese, E.~D., \& LaRoque, S.~J.\  2004, ApJ, 614, 56 

\bibitem[{Cowan}(1998)]{Cowan} Cowan, G. 1998, Statistical Data Analysis, Oxford.

\bibitem[{Fabian, Cowie \& Grindlay}(1981)]{fabian_depro} Fabian, A. C., Hu, E. M., Cowie, L. L., \& Grindlay, J. 1981, ApJ, 248, 47

\bibitem[{Fabian et al.}(2003)]{fabian} Fabian, A. C. et al. 2003, MNRAS 344, 43.

\bibitem[{Gilks, Richardson \& Spiegelhalter}(2003)]{mcmc} Gilks, W.~R. and Richardson, S. \& Spiegelhalter, D.~J., 1996, {Markov-Chain Monte-Carlo In Practice}. Chapman and Hall, Cambridge

\bibitem[{{Hastings}(1970)}]{ST/Has70} {Hastings}, W.~K. 1970, Biometrika, 57, 97

\bibitem[{Jernigan \& Vezie}(1996)]{jernigan} Jernigan, J. G. \& Vezie,
  M. 1996, in ASP Conf. Ser. 101, Astronomical Data Analysis Software and Systems V, ed. G. H. Jacoby \& J. Barnes (San Francisco: ASP), 167

\bibitem[{Kaastra et al.}(2003)]{kaastra} Kaastra, J. S.  et al. 2003, A \& A 413, 414.

\bibitem[{Kaastra}(1992)]{mekal3} Kaastra, J. S.  1992, An X-Ray Spectral Code for Optically Thin Plasmas (Internal SRON-Leiden Report, updated version 2.0)

\bibitem[{Lewis \& Bridle}(2002)]{lewis+bridle}  Lewis, A. \& Bridle S.L., 2002, {Phys.Rev.\ D} 66, 103511

\bibitem[{Liedahl, Osterheld, \& Goldstein}(1995)]{mekal4} Liedahl, D. A., Osterheld, A. L., \& Goldstein, W. H.  1995, \apj, 438, L115

\bibitem[{McKay}(2003)]{mckay} McKay, D.J.C., 2003,  {Information Theory, Inference, and Learning Algorithms} Cambridge University Press

\bibitem[{Markevitch et al.}(2000)]{markevitch} Markevitch, M. et al., 2000, ApJ 541, 542

\bibitem[{Marshall et al.}(2002)]{marshall1} Marshall, P. J., Hobson, M. P., Gull, S. F, Bridle, S. L. 2002, MNRAS 335, 1037.

\bibitem[{Marshall et al.}(2003)]{marshall2} Marshall, P. J., Hobson, M. P., Slosar, A. 2003, MNRAS 346, 489.

\bibitem[{{Metropolis} {et~al.}(1953){Metropolis}, {Rosenbluth}, {Rosenbluth}, {Teller}, \& {Teller}}]{ST/Met++53} {Metropolis}, N., {Rosenbluth}, A.~W., {Rosenbluth}, M.~N., {Teller}, A.~H., \& {Teller}, E. 1953, J.~Chem.~Phys., 21, 1087

\bibitem[{Mewe, Gronenschild, \& van den Oord}(1985)]{mekal1} Mewe, R., Gronenschild, E. H. B. M., \& van den Oord, G. H. J. 1985, \aaps, 62, 197

\bibitem[{Mewe, Lemen, \& van den Oord}(1986)]{mekal2} Mewe, R., Lemen, J. R., \& van den Oord, G. H. J.  1986, \aaps, 65, 511

\bibitem[{Morrison \& McCammon}(1983)]{wabs} Morrison, R. \& McCammon, D.  1983, \apj, 270, 119

\bibitem[{Skilling}(1998)]{massinf} Skilling, J., 1998,  in Maximum Entropy and Bayesian Methods, G.Erickson, J.T.Rychert \& C.Ray.Smith (eds)  Kluwer Academic Publishers, Dordtrecht

\bibitem[{\'O'Ruanaidh \& Fitzgerald}(1996)]{oruanaidh} {\'O'Ruanaidh} J.~J.~K.,  {Fitzgerald} W.~J.,  1996,  {Numerical Bayesian Methods Applied to Signal Processing}. Springer-Verlag, New-York

\bibitem[{Peterson et al.}(2001)]{peterson1} Peterson, J. R. et al. 2001, A \& A 365, 104.

\bibitem[{Peterson et al.}(2003)]{peterson2} Peterson, J. R. et al. 2003, ApJ 590, 207.

\bibitem[{Peterson, Jernigan, \& Kahn}(2004)]{peterson3} Peterson, J. R.,
  Jernigan, J. G., \& Kahn, S. M. 2004, ApJ 615, 545.

\bibitem[{Peterson et al.}(2006)]{peterson4} Peterson, J. R. et al. 2006, in preparation.

\bibitem[{Roberts, Gelman, \& Gilks}(1997)]{roberts} Roberts, G. O., Gelman, A., \& W. R. Gilks, Annals of Applied Probability 7, 110.

\bibitem[{Sanders et al.}(2004)]{sanders} Sanders, J. S., Fabian, A. C., Allen, S. W., \& Schmidt, R. W. 2004, MNRAS 349, 952.

\bibitem[Sanderson et al.\ (2004)]{sanderson} Sanderson, A. J., Finoguenov,
  A., Mohr, J. J. 2004, ApJ Accepted.

\bibitem[{Tegmark et al.}(2004)]{tegmark} Tegmark, M. et al., 2004, {Phys.Rev.\ D} 69, 103501.

\end{thebibliography}
\end{document}